\documentclass[pdflatex,sn-mathphys-num]{sn-jnl}
\usepackage{graphicx}%
\usepackage{multirow}%
\usepackage{amsmath,amssymb,amsfonts}%
\usepackage{amsthm}%
\usepackage{mathrsfs}%
\usepackage[title]{appendix}%
\usepackage{xcolor}%
\usepackage{textcomp}%
\usepackage{manyfoot}%
\usepackage{booktabs}%
\usepackage{algorithm}%
\usepackage{algorithmicx}%
\usepackage{algpseudocode}%
\usepackage{listings}%
\theoremstyle{thmstyleone}%
\theoremstyle{thmstyletwo}%
\theoremstyle{thmstylethree}%
\begin{document}

\title[Article Title]{Chaotic dynamics of charged particles near weakly magnetized black holes in Einstein-ModMax Theory}

\author[1]{\sur{Zijian Liu}}\email{m440123120@sues.edu.cn}

\author*[2]{\sur{Wenfu Cao}}\email{202411100001@stu.ujn.edu.cn}
\equalcont{These authors contributed equally to this work.}

\affil[1]{\orgdiv{School of Mathematics, Physics and Statistics}, \orgname{Shanghai University of Engineering Science},   \city{Songjiang}, \postcode{201620}, \country{People's Republic of China}}

\affil*[2]{\orgdiv{School of Physics and Technology}, \orgname{University of Jinan}, \city{Jinan}, \postcode{250022}, \country{People's Republic of China}}

\abstract{This paper presents a systematic study of the chaotic dynamics of charged test particles around purely magnetically charged black holes immersed in a uniform external magnetic field within the framework of Einstein-ModMax theory. By constructing an explicit symplectic integrator, we obtain high-precision numerical solutions of the equations of motion.
Combining the observational constraints from the Event Horizon Telescope (EHT) shadow images, we further restrict the parameter ranges of the model.
We apply Shannon entropy and MIPP (mutual information for particle pairs) as effective indicators to identify the chaotic behavior of charged test particles in the spacetime of this black hole.
Numerical results indicate that these indicators  can clearly distinguish between regular and chaotic motion of orbits in strong gravitational field systems.
Further analysis reveals that, compared to the key conserved quantities that determine the global dynamical behavior of the system---energy $E$ and angular momentum $L$, the sensitivity of the system parameters $e^{-\nu}$ and $Q_{m}$ to transitions in the orbital dynamical states is significantly reduced. This study provides a new perspective for a deeper understanding of the characterization and evolution mechanisms of chaotic dynamics in strong gravitational fields.}


\maketitle

\section{Introduction}\label{sec1}

Black holes are among the most fascinating objects in modern astrophysics, with their existence supported by a growing body of observational evidence.
This is evidenced by gravitational-wave signals from binary black hole mergers captured by the Laser Interferometer Gravitational-Wave Observatory (LIGO) \cite{LIGOScientific:2016aoc}, as well as  the shadow images of M87* and the Milky Way's center Sgr A* obtained by the EHT \cite{EventHorizonTelescope:2019dse, EventHorizonTelescope:2022wkp}.
These discoveries not only validate the predictions of general relativity in strong gravitational fields but also establish a crucial testing ground for a range of modified gravity models \cite{Nashed:2019tuk, Brihaye:2019kvj,Konoplya:2019fpy,Long:2020wqj,Ajith:2020ydz,Sau:2022afl,Hu:2023pyd}.

Observational evidence reveals the ubiquitous presence of magnetic fields around astrophysical black holes \cite{Eatough:2013nva,EventHorizonTelescope:2021srq}.
These magnetic fields originate from the dynamics of ionized matter and plasma within the accretion disks surrounding black holes, which may facilitate energy transfer from the disk to relativistic jets \cite{Stuchlik:2015nlt}.
The external magnetic field surrounding the black hole satisfies the vacuum Maxwell equations, and one well-known approach to solving these is the Wald solution \cite{Wald:1974np}.
However, this solution has well-defined limits of applicability: it requires the black hole to be electrically neutral (Kerr or Schwarzschild black holes) and the external magnetic field to be weak, with its strength significantly below  $10^{19}M_{\bigodot}/M$  Gauss, where  $M_{\bigodot}$  and  \emph{M}  denote the solar mass and black hole mass, respectively \cite{Frolov:2010mi, Lee:2022rtg}.
When the applicability conditions of the Wald solution are violated, such as in Kerr-Newman black holes \cite{Newman:1965my} or Kerr-Newman-Melvin spacetimes \cite{Ernst:1976bsr}, these configurations fail to satisfy the source-free Maxwell equations.
In this case, the Wald vector potential must be appropriately modified and generalized \cite{Abdujabbarov:2009az,Abdujabbarov:2011uc,Azreg-Ainou:2016tkt,Cao:2024ihv,Siahaan:2024ioa,Zhang:2025qia}. When a magnetic field is present in curved spacetime, it leads to the non-integrability of the Hamiltonian system associated with the equations of motion for charged particles, resulting in chaotic behavior \cite{Karas:1992, Li:2018wtz}.

In recent years, the Einstein-ModMax weakly magnetized black hole theory has received significant attention \cite{Amirabi:2020mzv,Barrientos:2022bzm,Siahaan:2024ioa}.
For the dual-charge black hole with both electric and magnetic charges, this theory applies the Wald magnetization method to the Einstein gravity theory with nonlinear electrodynamics, constructing the Einstein-ModMax magnetized dual-charge black hole solution, and analyzes the motion characteristics of charged test particles in this background.
However, under the Einstein-ModMax theoretical framework, research remains limited regarding the impact of purely magnetically charged black holes on the chaotic motion of charged test particles.
Motivated by this, the present paper explores the chaotic dynamical characteristics of charged test particles around a purely magnetically charged black hole immersed in an external magnetic field.

Numerical methods play an important role in analyzing the non-integrability of Hamiltonian systems.
Explicit symplectic integrators are algorithms specifically designed for integrating the equations of motion of such systems.
By preserving the system's symplectic structure, these integrators effectively prevent long-term numerical drift in conserved quantities---such as energy $E$ and angular momentum $L$---thereby yielding more accurate and stable simulation results.
High-precision symplectic integrators can be constructed by splitting the Hamiltonian into subcomponents, employing specific integration schemes, and optimizing algorithm parameters \cite{Wang:2021gja, Wang:2021xww, Wang:2021yqk, Wu:2021rrd, Wu:2022nye, Zhou:2022uht, Liu:2023mdu, Wu:2024ehd, Lu:2024srb}. 
In addition to reliable symplectic integrators, efficient chaos indicators are also required to distinguish between regular and chaotic dynamics.
The Poincar$\acute{e}$ section is a classical tool for analyzing chaotic behavior; by examining the distribution of points on the section, one can extract key information regarding the system's periodicity, stability, and chaotic nature.
Furthermore, the maximum Lyapunov exponent is a widely used diagnostic.
When this exponent is positive, neighboring trajectories in phase space diverge exponentially over time, signifying chaos \cite{Wu:2003pe}.
The fast Lyapunov indicator, a generalized version of the Lyapunov exponent, facilitates more rapid detection of chaotic behavior \cite{Wu:2006rx}.
Complementing these classical approaches, recent studies have demonstrated that Shannon entropy and MIPP are effective indicators.
Specifically, Shannon entropy successfully distinguishes between periodic, quasi-periodic, and chaotic orbits; as a system transitions from regular to chaotic regimes, the entropy value increases, and its fluctuations become significantly more pronounced \cite{Cao:2024bjk}. 
Similarly, MIPP enables the direct identification of transitions between regular and chaotic orbital motion through scanning diagrams, where values close to 1 correspond to regular orbits and values close to 0 indicate chaotic behavior. 
Moreover, it provides higher computational efficiency than the fast Lyapunov indicator \cite{Cao:2024rvo}.

The paper is organized as follows. In Sec. \ref{sec:two}, the metric of the magnetically charged black hole  and the electromagnetic four-vector are introduced. In Sec. \ref{sec:three}, the dynamical characteristics of charged test particles are discussed. Finally, in Sec. \ref{sec:four}, our main results are summarized.

\section{Metric of the magnetic charge black hole and electromagnetic four-vector}\label{sec:two}

To investigate the chaotic dynamics of charged test particles near a purely magnetically charged Einstein-ModMax black hole immersed in a uniform external magnetic field, it is essential to first specify the spacetime structure of the black hole and the configuration of its electromagnetic field.

\subsection{Metric of the magnetic charge black hole}\label{subsec1}

This work focuses on purely magnetically charged black holes immersed in a uniform external magnetic field. The spacetime metric of such a black hole is described by
\begin{eqnarray}
-d\tau^2 =ds^{2} &=& g_{\alpha\beta}dx^{\alpha}dx^{\beta} \nonumber \\
&=& -f(r)dt^{2} \nonumber +\frac{1}{f(r)} dr^{2}\nonumber \\
& &+r^{2}d \theta^{2} +r^{2}\sin^{2} \theta d \varphi^2,
\end{eqnarray}
with
\begin{eqnarray}
f(r)=(1-\frac{2M}{r}+\frac{e^{-\nu}Q_{m}^{2}}{r^{2}}),
\end{eqnarray}
where $Q_{m}$  is the magnetic charge of the black hole, and  $\nu$  (with  $\nu\geq0$) is the nonlinear parameter, which achieves natural screening of the black hole's electromagnetic charge.
In this paper,  $e^{-\nu}$  is considered as a combined parameter.
The event horizon of the black hole is determined by  $f(r)=0$, and the solution to the equation is  $r=M\pm\sqrt{M^{2}-e^{-\nu}Q_{m}^2}$.
For convenience, this paper adopts geometric units, by setting both the black hole mass  $M$  and the test particle mass  $m$  to 1, i.e.,  $M=m=1$.

The system described by this metric follows a Lagrangian framework:
\begin{eqnarray}
\ell=\frac{1}{2}(\frac{ds}{d\tau})^{2}=\frac{1}{2}g_{\mu\nu}\dot{x}^{\mu}\dot{x}^{\nu}.
\end{eqnarray}
Four-velocity components in spacetime satisfying the normalization condition:
\begin{eqnarray}
\mathbf{U}\cdot\mathbf{U}= g_{\mu\nu}U_{\mu}U^{\mu}=g_{\mu\nu}\dot{x}^{\mu}\dot{x}^{\nu}=-1.
\end{eqnarray}
The covariant generalized momentum derived from the Lagrangian is:
\begin{eqnarray}
p_{\mu}=\frac{\partial\ell}{\partial\dot{x}^{\mu}}=g_{\mu\nu}\dot{x}^{\nu},
\end{eqnarray}
the Lagrangian does not depend explicitly on $t$ and $\varphi$, so there are two conserved momentum components
\begin{eqnarray}
p_{t}=g_{tt}\dot{t}=-(1-\frac{2}{r}+\frac{e^{-\nu}Q_{m}^{2}}{r^{2}})\dot{t}=-E,
\end{eqnarray}
\begin{eqnarray}
p_{\varphi}=g_{\varphi\varphi}\dot{\varphi}=r^{2}\sin^{2}\theta\dot{\varphi}=L,
\end{eqnarray}
where $E$ denotes the energy of a test particle orbiting the rotating body, while $L$ represents its angular momentum.
In the same way as in classical mechanics, this Lagrangian is precisely equivalent to the Hamiltonian:
\begin{eqnarray}
\mathcal{H} &=& \frac{1}{2}g^{\mu\nu}p_{\mu}p_{\nu} \nonumber \\
&=& F(r,\theta)+\frac{1}{2}(1-\frac{2}{r}+\frac{e^{-\nu}Q_{m}^{2}}{r^{2}})p_{r}^{2}+\frac{1}{2r^{2}}p_{\theta}^{2},
\end{eqnarray}
the function   $F(r, \theta)$  includes the contributions from energy  $E$  and angular momentum $L$
\begin{eqnarray}
F(r,\theta) &=& \frac{1}{2}(g^{tt}E^{2}+g^{\varphi\varphi}L^{2})\nonumber \\
&=& -\frac{1}{2}(1-\frac{2}{r}+\frac{e^{-\nu}Q_{m}^{2}}{r^{2}})^{-1}E^{2}+\frac{1}{2r^{2}\sin^{2}\theta}L^{2},
\end{eqnarray}
since the four-velocity $\dot{x}^\mu$ always satisfies the  Eq. (4), the Hamiltonian  is  equal to a known constant
\begin{eqnarray}
\mathcal{H}&=&-\frac{1}{2}.
\end{eqnarray}
This constant relation provides an integral constraint for the equations of motion.

A fourth constant of motion exists, which is derived from the separation of variables in the Hamilton-Jacobi equation of the Hamiltonian system. The expression for the fourth constant is \cite{Carter:1968rr}:
\begin{eqnarray}
C_{k}&=&(1-\frac{2}{r}+\frac{e^{-\nu}Q_{m}^{2}}{r^{2}})^{-1}r^{2}E^{2} \nonumber \\
& &-r^{2}-r^{2}p_{r}^{2}(1-\frac{2}{r}+\frac{e^{-\nu}Q_{m}^{2}}{r^{2}}) \nonumber \\
&=& p_{\theta}^{2}+\frac{L^{2}}{\sin^{2} \theta}.
\end{eqnarray}
The four constants of motion $(H, E, L, C_{k})$ render Eq. (8) completely solvable.

\subsection{Electromagnetic four-vector}\label{subsec2}

An asymptotically uniform magnetic field is an important subject of study in astrophysical environments. Wald \cite{Wald:1974np} studied the analytic solution for the electromagnetic field of a static axisymmetric black hole in a uniform magnetic field. The core idea is to use the Killing symmetries of the spacetime to construct the electromagnetic four-vector.  The electromagnetic four-vector  $A^\alpha$ is constructed from a linear combination of the timelike Killing vector $\xi_{(t)}^\alpha = (1,0,0,0)$ and the spacelike Killing vector $\xi_{(\varphi)}^\alpha = (0,0,0,1)$.
\begin{eqnarray}
A^{\alpha}=C_{1}\xi_{(t)}^{\alpha}+C_{2}\xi_{(\varphi)}^{\alpha},
\end{eqnarray}
for uncharged black holes, set $C_1 = 0$ and $C_2 = \dfrac{B}{2} ( 1 - \dfrac{e^{-\nu} Q_m^2}{r^2} )$ \cite{Siahaan:2024ioa}. It is evident that the four-vector $A^\alpha$ has a non-zero covariant component:
\begin{eqnarray}
A_{\varphi}=\frac{B}{2}(1-\frac{e^{-\nu}Q_{m}^{2}}{r^{2}})g_{\varphi\varphi}=\frac{B}{2}(r^{2}-e^{-\nu}Q_{m}^{2})\sin^{2}\theta.
\end{eqnarray}

\section{Motion of a charged particle in an external magnetic field around a magnetic black hole}\label{sec:three}

To address the non-integrability caused by the external magnetic field, an explicit symplectic integrator is employed to obtain high-precision numerical solutions. Shannon entropy and MIPP indicators are applied to identify chaotic particle motion, using the parameter space constrained by EHT observations.

\subsection{Constructing an explicit symplectic integrator}\label{subsec1}
Symplectic integrators are a class of numerical methods specifically designed for solving Hamiltonian systems.
Their core advantage lies in preserving the symplectic structure of the system, thereby demonstrating outstanding numerical stability in long-term simulations.
In general, explicit symplectic integrators outperform implicit ones for the same order of accuracy in terms of computational efficiency.
The key to constructing an explicit symplectic integrator is that the Hamiltonian must be decomposable into several analytically solvable sub-Hamiltonians \cite{Wang:2021gja}.

The motion of a charged particle  $q$  in an external magnetic field can be described by a Hamiltonian system:
\begin{eqnarray}
K &=& \frac{1}{2}g^{\mu\nu}(p_{\mu}-qA_{\mu})(p_{\nu}-qA_{\nu}) \nonumber \\
&=& -\frac{1}{2}(1-\frac{2}{r}+\frac{e^{-\nu}Q_{m}^{2}}{r^{2}})^{-1}E^{2} \nonumber \\
& &+\frac{1}{2}(1-\frac{2}{r}+\frac{e^{-\nu}Q_{m}^{2}}{r^{2}})p_{r}^{2}+\frac{p_{\theta}^{2}}{2r^{2}} \nonumber \\
& &+\frac{1}{2r^{2}\sin^{2}\theta}[L-\frac{1}{2}\beta(r^{2}-e^{-\nu}Q_{m}^{2})\sin^{2}\theta]^{2},
\end{eqnarray}
where $\beta = qB$, and the Hamiltonian  $K$  also satisfies the constraint condition  $K = -\frac{1}{2}$.

The Hamiltonian Eq. (14) can be decomposed into the following form
\begin{eqnarray}
K =K_{1}+K_{2}+K_{3}+K_{4}+K_{5}.
\end{eqnarray}
The sub-Hamiltonians are given by
\begin{eqnarray}
K_{1}(r,\theta) &=& \frac{1}{2r^{2}\sin^{2}\theta}[L-\frac{1}{2}\beta(r^{2}-e^{-\nu}Q_{m}^{2})\sin^{2}\theta]^{2} \nonumber \\
& & -\frac{1}{2}(1-\frac{2}{r}+\frac{e^{-\nu}Q_{m}^{2}}{r^{2}})^{-1}E^{2} ,\\
K_{2}&=&\frac{1}{2}p_{r}^{2} , \\
K_{3}&=&-\frac{1}{r}p_{r}^{2} , \\
K_{4}&=&\frac{e^{-\nu}Q_{m}^{2}}{2r^{2}}p_{r}^{2} , \\
K_{5}&=&\frac{1}{2r^{2}}p_{\theta}^{2} .
\end{eqnarray}
The regular equation for the sub-Hamiltonian  $K_{1}$  is
\begin{eqnarray}
\frac{dp_{r}}{d\tau}&=&-\frac{\partial K_{1}(r,\theta)}{\partial r},\nonumber \\
\frac{dp_{\theta}}{d\tau}&=&-\frac{\partial K_{1}(r,\theta)}{\partial \theta}.
\end{eqnarray}
The regular equations for the other sub-Hamiltonians can be written as:
\begin{eqnarray}
K_{2}: \frac{dr}{d\tau}&=&p_{r} , \dot{p}=0 , \\
K_{3}: \frac{dr}{d\tau}&=&-\frac{2}{r}p_{r} , \frac{dp_{r}}{d\tau}=-\frac{1}{r^{2}}p_{r}^{2} ,\\
K_{4}: \frac{dr}{d\tau}&=&\frac{e^{-\nu}Q_{m}^{2}}{r^{2}}p_{r} , \frac{dp_{r}}{d\tau}=\frac{e^{-\nu}Q_{m}^{2}}{r^{3}}p_{r}^{2} , \\
K_{5}: \frac{d\theta}{d\tau}&=&\frac{1}{r^{2}}p_{\theta} , \frac{dp_{r}}{d\tau}=\frac{1}{r^{3}}p_{\theta}^{2} , \dot{r}=\dot{p}=0.
\end{eqnarray}
Each equation is solved independently. Let $(r_0$, $\theta_0$, $p_{r0}$, $p_{\theta0})$ denote the initial values at proper time $\tau_0$, and $(r$, $\theta$, $p_r$, $p_\theta)$ represent the analytical solution at time $\tau$ ($\tau=\tau_0+h$). The evolution operator for $K$ is denoted by $\widetilde{K}$.
\begin{eqnarray}
\widetilde{K}_{1}: p_{r}(\tau) &=& p_{r0}-\tau\frac{\partial K_{1}(r,\theta)}{\partial r}|_{(r_{0},\theta_{0})} , \nonumber \\
 p_{\theta}(\tau) &=& p_{\theta0}-\tau\frac{\partial K_{1}(r,\theta)}{\partial \theta}|_{(r_{0},\theta_{0})},\\
\widetilde{K}_{2}: r(\tau)&=& r_{0}+\tau p_{r0},\\
\widetilde{K}_{3}: r(\tau)&=& [\frac{(r_{0}^{2}-3\tau p_{r0})^{2}}{r_{0}}]^{\frac{1}{3}} ,\nonumber \\
p_{r}(\tau)&=& p_{r0}(\frac{r_{0}^{2}-3\tau p_{r0}}{r_{0}^{2}})^{\frac{1}{3}}, \\
\widetilde{K}_{4}: r(\tau)&=& (\frac{2\tau e^{-\nu}Q_{m}^{2}p_{r0}}{r_{0}}+r_{0}^{2})^{\frac{1}{2}} ,\nonumber \\
p_{r}(\tau)&=& \frac{p_{r0}}{r_{0}}(\frac{2\tau e^{-\nu}Q_{m}^{2}p_{r0}}{r_{0}}+r_{0}^{2})^{\frac{1}{2}}, \\
\widetilde{K}_{5}: \theta(\tau)&=&\theta_{0}+\frac{\tau p_{\theta0}}{r_{0}^{2}} , \nonumber \\
p_{r}(\tau)&=& p_{r0}+\frac{\tau p_{\theta0}^{2}}{r_{0}^{3}}.
\end{eqnarray}
These five sub-Hamiltonians have explicit analytical solutions with respect to the proper time $\tau$. Their solutions correspond to the evolution operators $\widetilde{K}_1^\mathcal{H}$, $\widetilde{K}_2^\mathcal{H}$, $\widetilde{K}_3^\mathcal{H}$, $\widetilde{K}_4^\mathcal{H}$, $\widetilde{K}_5^\mathcal{H}$. These evolution operators can be combined to form a second-order explicit symplectic integrator for $K$. By setting the time step size as  $h$, we obtain a second-order explicit symplectic integrator:
\begin{eqnarray}
S_{2}^{K}(h)&=&\kappa^{*}\cdot\kappa(h) \nonumber \\
&=& \widetilde{K}_5^{\mathcal{H}}(\frac{h}{2})\cdot \widetilde{K}_4^{\mathcal{H}}(\frac{h}{2})\cdot \widetilde{K}_3^{\mathcal{H}}(\frac{h}{2})\cdot \widetilde{K}_2^{\mathcal{H}}(\frac{h}{2})\cdot \widetilde{K}_1^{\mathcal{H}}(h) \nonumber \\
& & \cdot \widetilde{K}_2^{\mathcal{H}}(\frac{h}{2})\cdot \widetilde{K}_3^{\mathcal{H}}(\frac{h}{2})\cdot \widetilde{K}_4^{\mathcal{H}}(\frac{h}{2}) \cdot \widetilde{K}_5^{\mathcal{H}}(\frac{h}{2}),
\end{eqnarray}
where two first-order solvers are
\begin{eqnarray}
\kappa^{*}(h)&=& \widetilde{K}_5^{\mathcal{H}}(\frac{h}{2})\cdot \widetilde{K}_4^{\mathcal{H}}(\frac{h}{2})\cdot \widetilde{K}_3^{\mathcal{H}}(\frac{h}{2})\nonumber \\
& &\cdot \widetilde{K}_2^{\mathcal{H}}(\frac{h}{2})\cdot \widetilde{K}_1^{\mathcal{H}}(\frac{h}{2}) ,
\end{eqnarray}
\begin{eqnarray}
\kappa(h)&=& \widetilde{K}_1^{\mathcal{H}}(\frac{h}{2})\cdot \widetilde{K}_2^{\mathcal{H}}(\frac{h}{2})\cdot \widetilde{K}_3^{\mathcal{H}}(\frac{h}{2})\nonumber \\
& &\cdot \widetilde{K}_4^{\mathcal{H}}(\frac{h}{2})\cdot \widetilde{K}_5^{\mathcal{H}}(\frac{h}{2}) .
\end{eqnarray}
By symmetrically combining three second-order methods, a fourth-order explicit symplectic scheme can be obtained \cite{Yoshida:1990zz}:
\begin{eqnarray}
S_{4}^{K}(h)&=&S_{2}^{K}(\gamma h)\cdot S_{2}^{K}(\delta h)\cdot S_{2}^{K}(\gamma h),
\end{eqnarray}
the combination coefficients must satisfy
\begin{eqnarray}
\delta+2\gamma=1, \delta^{3}+2\gamma^{3}=0,
\end{eqnarray}
where $\gamma=\frac{1}{2-\sqrt[3]{2}}$ , $\delta=1-2\gamma$. By adding more components of the first-order operators  $\kappa^{*}$  and  $\kappa$, an optimized fourth-order partitioned Runge-Kutta (PRK) symplectic algorithm can be obtained \cite{Zhou:2022uht}:
\begin{eqnarray}
PRK_{6}4 &=& \kappa^{*}(\alpha_{12}h)\cdot \kappa(\alpha_{11}h)\cdot\cdot\cdot \kappa^{*}(\alpha_{2}h)\cdot \kappa(\alpha_{1}h),
\end{eqnarray}
the time coefficients are
\begin{eqnarray}
\alpha_{1}&=& \alpha_{12}=0.079203696431196 ,\nonumber \\
\alpha_{2}&=& \alpha_{11}=0.130311410182166 ,\nonumber \\
\alpha_{3}&=& \alpha_{10}=0.222861495867608 ,\nonumber \\
\alpha_{4}&=& \alpha_{9}=-0.366713269047426 ,\nonumber \\
\alpha_{5}&=& \alpha_{8}=0.324648188689706 ,\nonumber \\
\alpha_{6}&=& \alpha_{7}=0.109688477876750 .\nonumber
\end{eqnarray}

\begin{figure}[H]
\centering
\includegraphics[width=0.32\textwidth]{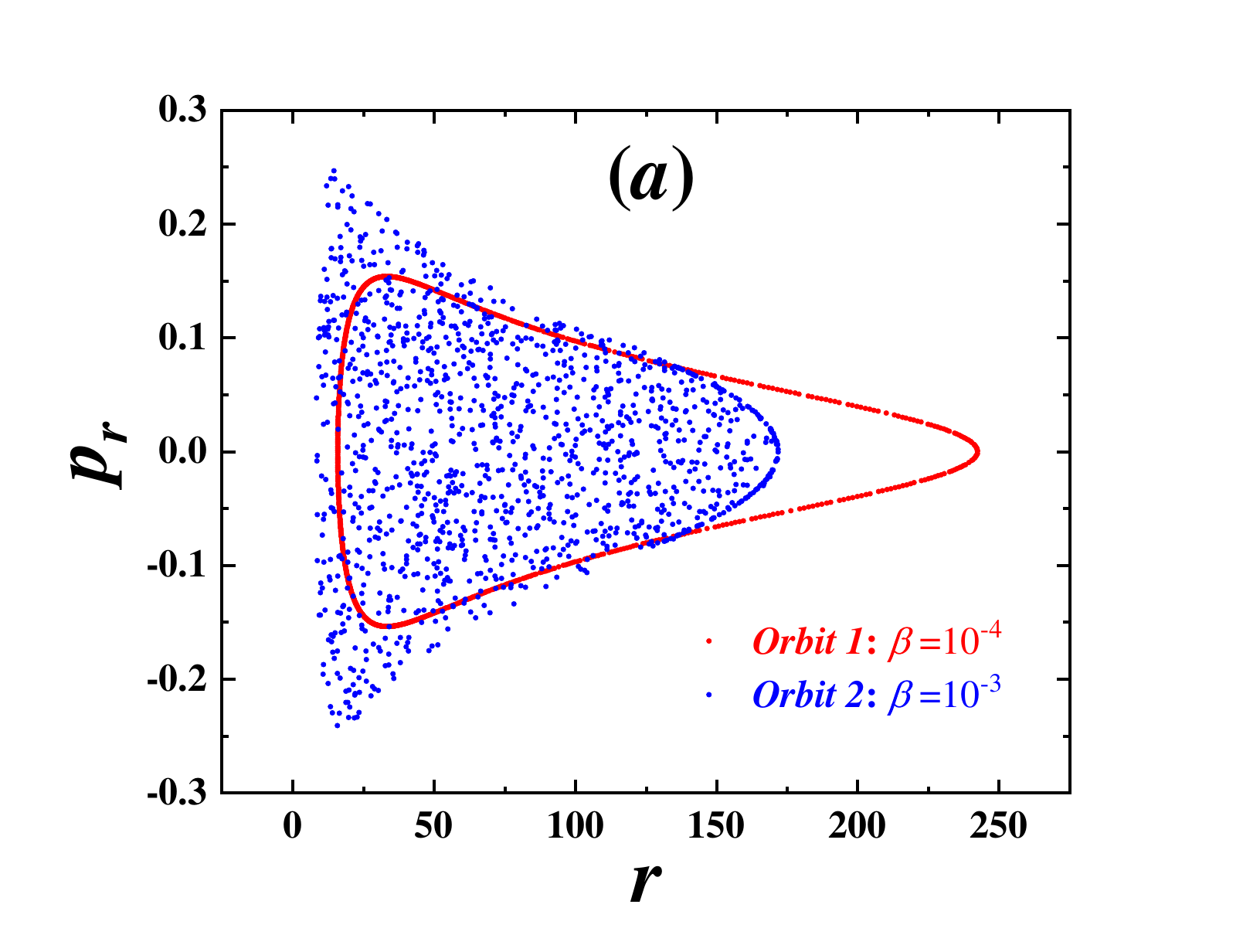}
\includegraphics[width=0.32\textwidth]{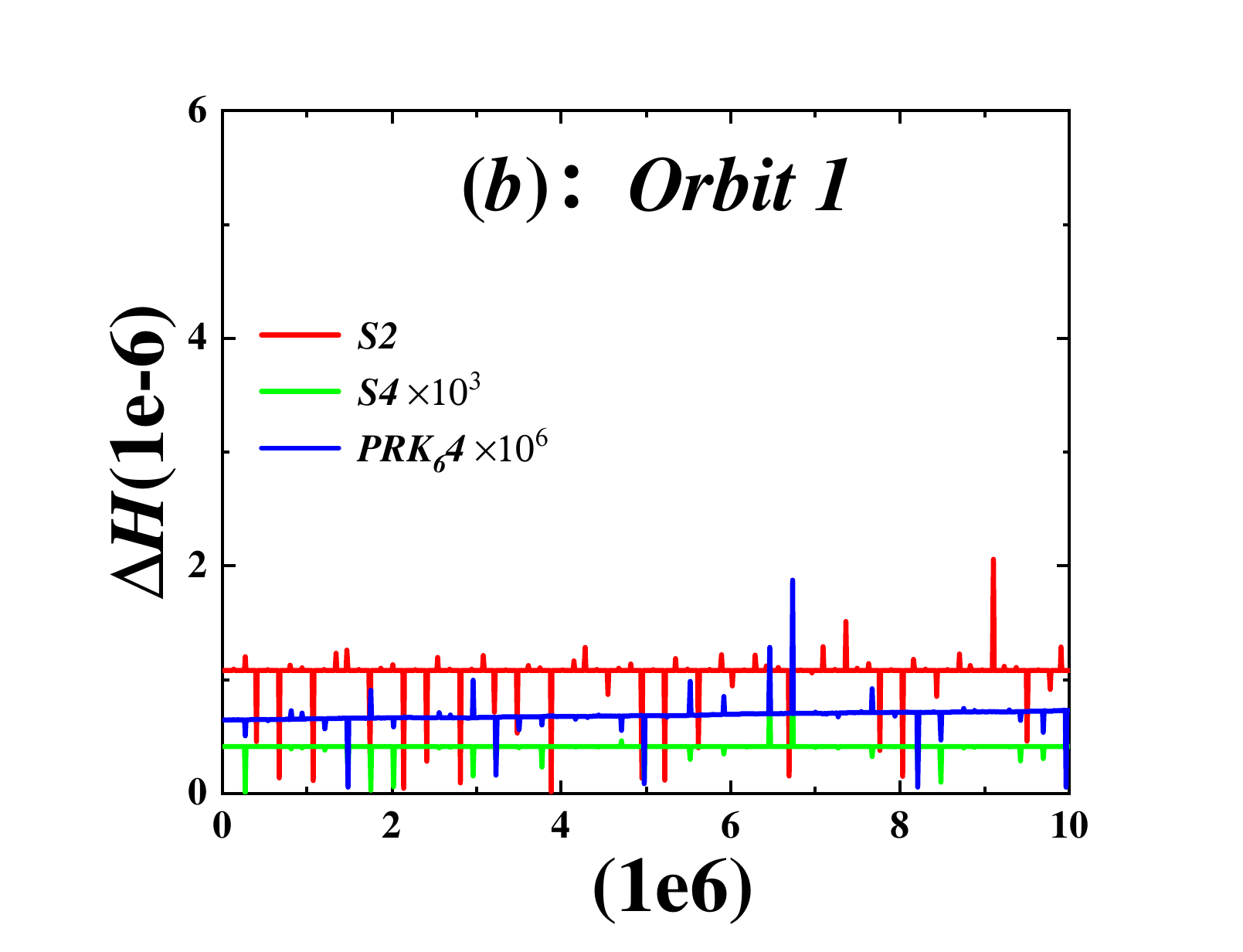}
\includegraphics[width=0.32\textwidth]{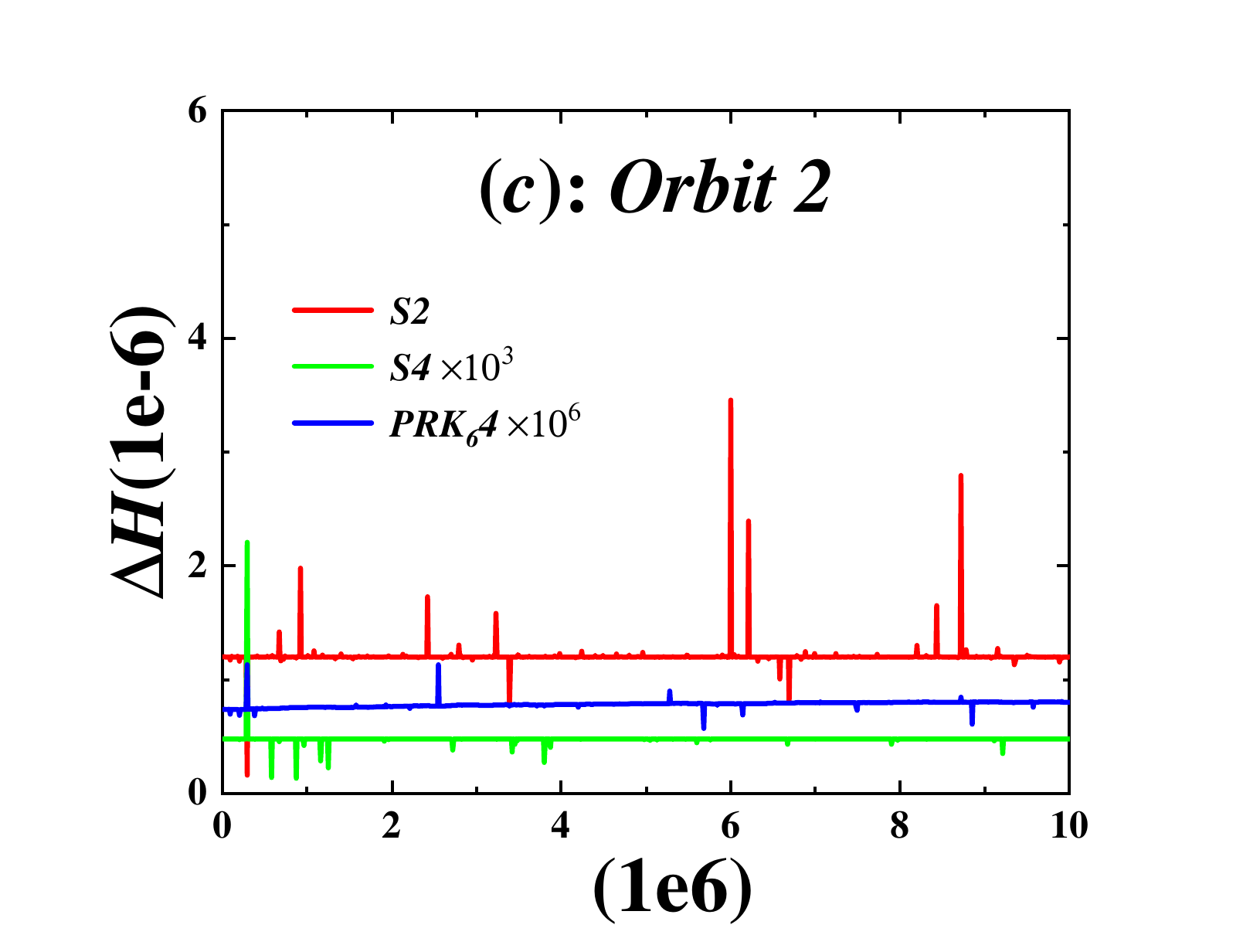}

\caption{ (a): Poincar$\acute{e}$ sections for Orbits 1 and 2.
(b)-(c): Hamiltonian errors $\Delta{\mathcal{H}}=\mathcal{H}+\frac{1}{2}$
for several explicit symplectic integrators for Orbits 1 and 2.
The two orbits share common parameters $E = 0.996$, $L = 4.6$, $Q_{m}=1$, $e^{-\nu}=0.24$ and initial conditions $r = 16$, $\theta=\frac{\pi}{2}$.
The magnetic field parameter is $\beta=10^{-4}$ for Orbit 1 and $\beta=10^{-3}$ for Orbit 2.}\label{fig1}
\end{figure}

To assess the performance of these symplectic integrators, we perform numerical simulations for representative orbits.
We set the integration step $h=1$, and the physical parameters as $E=0.996$, $L=4.6$, $Q_{m}=1$ and $e^{-\nu}=0.24$.
The initial conditions are $r=16$, $\theta=\frac{\pi}{2}$ and $p_{r}=0$.
We consider two magnetic field strengths: $\beta=10^{-4}$(Orbit 1) and $\beta=10^{-3}$(Orbit 2).
As shown in Fig. 1, when the integration time reaches $10^{7}$, the three integrators $S_{2}$, $S_{4}$, and $PRK_{6}4$ exhibit no secular drift in the Hamiltonian error $\Delta{\mathcal{H}}=\mathcal{H}+\frac{1}{2}$ for the two representative orbits shown in Fig. 1(b)-(c).
This demonstrates the well-known advantage of symplectic integrators in long-time energy conservation.
It can also be seen that the accuracy of $S_{4}$ is about three orders of magnitude better than that of $S_{2}$, while $PRK_{6}4$ further improves the accuracy by approximately three orders of magnitude compared with $S_{4}$.
These results confirm the high numerical precision of the $PRK_{6}4$ scheme. Therefore, the $PRK_{6}4$ integrator is adopted in the subsequent computations.

\subsection{Parameter space constraints}\label{subsec2}
By utilizing the correspondence between the angular radius of the bright ring of Sgr $A^{*}$ measured by the EHT and its theoretical shadow angular radius, combined with the mass-to-distance ratio of Sgr $A^{*}$ determined from observations by Keck and the Very Large Telescope Interferometer (VLTI), constraints on the ratio of the black hole shadow radius to its mass can be derived \cite{Vagnozzi:2022moj}
\begin{eqnarray}
4.55\lesssim r_{sh} \lesssim5.22,
\end{eqnarray}
this range provides strict constraints on the parameter space of theoretical models, and the chaotic dynamics investigated here are restricted to the parameter region consistent with the observed shadow radius.

\begin{figure}[H]
\centering
\includegraphics[width=0.5\textwidth]{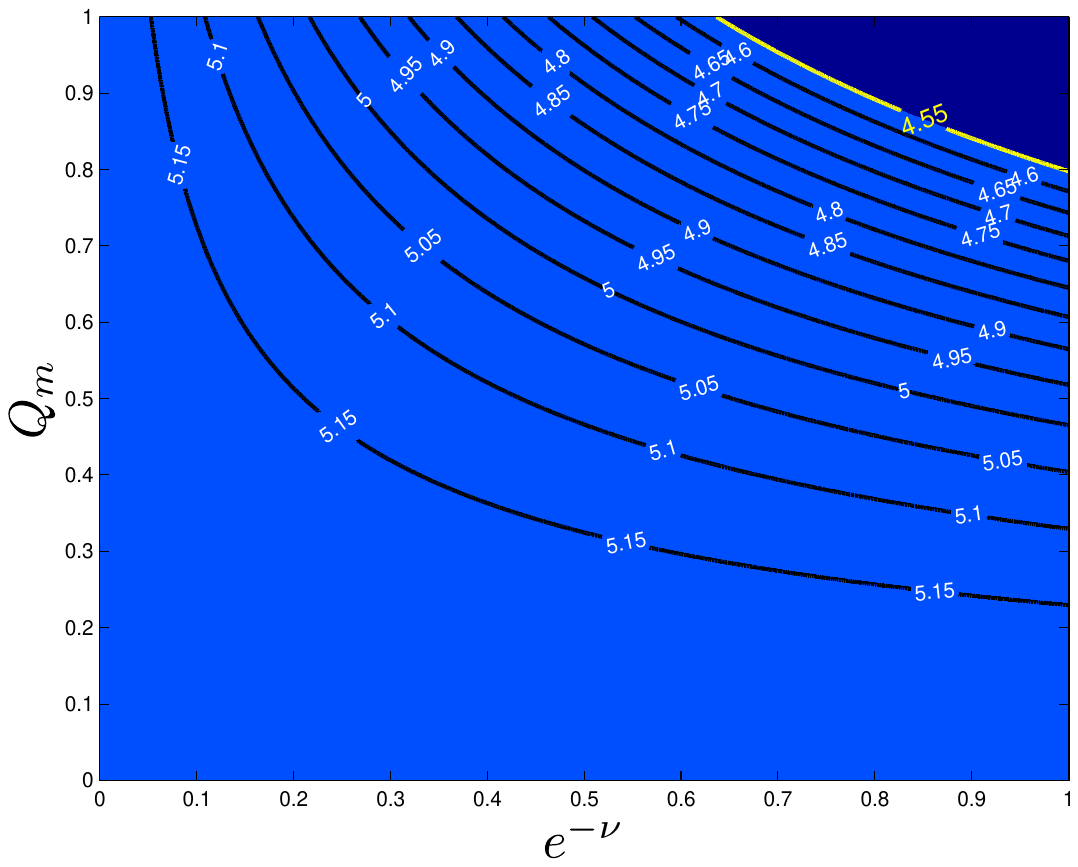}
\caption{Constraints on the Einstein-ModMax black hole parameters $(e^{-v},Q_{m})$ derived from the EHT shadow observation.
The contour lines represent the values of the shadow radius  $r_{sh}$.
The light blue region denotes the allowed parameter space consistent with the observational constraint $4.55\lesssim r_{sh} \lesssim5.22$,
whereas the dark blue region represents the parameter space excluded by these observations.
The boundary between the two regions corresponds to the critical values of $r_{sh}=4.55$.}\label{fig2}
\end{figure}

Next, we will use the constraints in Eq. (37) to limit the parameters of the model. The expression for the black hole shadow radius is
\begin{eqnarray}
r_{sh}&=&\frac{\sqrt{2}(3+\sqrt{9-8e^{-\nu}Q_{m}^{2}})}{\sqrt{4+\frac{\sqrt{9-8e^{-\nu}Q_{m}^{2}-3}}{e^{-\nu}Q_{m}^{2}}}}.
\end{eqnarray}
Combining Eqs. (37) and (38) yields the permissible ranges for the parameters  $e^{-\nu}$ and $Q_{m}$. Fig. 2 shows that when $e^{-\nu}=1$, $Q_{m}=0.8$; when $Q_{m}=1$, $e^{-\nu}=0.64$. When $e^{-\nu}$ takes values between 0 and 0.64, the allowed region for $Q_{m}$ in Fig. 2 is [0, 1]. Similarly, when  $Q_{m}$ takes values between 0 and 0.8, the allowed region for $e^{-\nu}$ in Fig. 2 is [0, 1]. However, when $e^{-\nu}$ takes values between 0.64 and 1, the allowed region for $Q_{m}$ in Fig. 2 lies below the yellow curve. Likewise, when $Q_{m}$ takes values between 0.8 and 1, the allowed region for  $e^{-\nu}$ in Fig. 2 lies below the yellow curve. This constraint ensures the astrophysical plausibility of the theoretical model, and all subsequent analyses will be confined to this region.

\subsection{Chaos detection method}\label{subsec3}
To accurately distinguish the regular and chaos motion of charged particles, this paper combines three chaos indicators, describing the dynamical characteristics from three dimensions, including phase space structure, probability distribution, and statistical correlation. These methods include the Poincar$\acute{e}$ section, Shannon entropy, and MIPP.

The Poincar$\acute{e}$ section is a classic tool for detecting chaos. Its core idea is to select a hyperplane in phase space and record the phase space points each time the system's trajectory crosses this plane.
A single point on the map represents a periodic orbit, while one or more closed curves indicate a quasi-periodic system. If these points are randomly distributed, the system is chaotic.

Shannon entropy, also known as information entropy, is a core concept in information theory, used to quantify the uncertainty or randomness of information \cite{Shannon:1948dpw}. Its definition can be expressed by the following formula:
\begin{eqnarray}
H(X)&=&-\sum_{i=1}^{n}p(x_{i})\log_{b}p(x_{i}),
\end{eqnarray}
where $H(X)$ is the Shannon entropy of the random variable  $X$, $p(x_{i})$ represents the probability of occurrence of event $x_{i}$, $n$ is the total number of all possible events.  $b$  is the base of the logarithm, typically set to 2.

The maximum value of Shannon entropy depends on the number of possible events  $n$  in the system. When each event has equal probability  $p(x_{i})=\frac{1}{n}$, the entropy reaches its maximum value, given by
\begin{eqnarray}
H_{max}(X)&=&\log_{b}n.
\end{eqnarray}
We generate  $10^{7}$ particle coordinate data $r$ through numerical integration of Hamiltonian equations of motion.
These data are divided  into 20 large intervals, with each large interval containing  $5\times10^{5}$  data points.
Next, we calculate the Shannon entropy for each large interval to observe changes in entropy. Taking the first large interval as an example, within this interval, we fix the range of  $r$  (which must include most of the particles to ensure accurate probability calculations. In this paper, the range is set to  $10\leq r\leq500$), and divide it into 100 small intervals, counting the number of particles that fall into each small interval to calculate the probability for each small interval. For example, in the  $i$-$th$ small interval, if there are  $y_{i}$  particles, then the probability for this interval is
\begin{eqnarray}
p(x_{i})&=&\frac{y_{i}}{5\times10^{5}}.
\end{eqnarray}
After calculating the probability for each small interval, the Shannon entropy for the first large interval can be obtained using the Shannon entropy Eq. (39) (if  $p(x)=0$, it is conventionally taken that  $0\log0=0$). The Shannon entropy for the remaining 19 intervals can be calculated using the same method and presented in the form of a line graph. The results show that the Shannon entropy of the chaotic system exhibits fluctuations, whereas that of a regular system remains relatively stable (a straight line).

Mutual Information (MI) \cite{Parrondo:2015thh} is an important concept in information theory, used to measure the statistical dependence between two random variables, i.e., how much information one variable contains about the other. If the two variables are independent, the mutual information is 0; if they are perfectly correlated, the mutual information reaches its maximum value. For two discrete random variables  $X$  and  $Y$,  $P(x,y)$  is their joint probability distribution, and  $P(x)$  and  $P(y)$  are the marginal probability distributions of  $X$  and  $Y$, respectively. Their mutual information  $I(X;Y)$  is defined as:
\begin{eqnarray}
I(X;Y)=\sum_{x\in X}\sum_{y\in Y}P(x,y)\log[\frac{P(x,y)}{P(x)P(y)}].
\end{eqnarray}
This equation can also be expressed as the relationship between entropy and joint entropy
\begin{eqnarray}
I(X;Y)= H(X)+H(Y)-H(X,Y) ,
\end{eqnarray}
where  $H(X)$  and  $H(Y)$  represent the Shannon entropy of the random variables  $X$  and  $Y$, respectively, and  $H(X,Y)$  is the joint Shannon entropy.

The numerical range of the original mutual information depends on the Shannon entropy of the variables. To characterize the orbital state transitions clearly, the mutual information is normalized as:
\begin{eqnarray}
\bar{I}(X;Y)=\frac{I(X;Y)}{H(X,Y)}.
\end{eqnarray}
$\bar{I}(X;Y)$ represents the normalized mutual information. It standardizes mutual information to the ratio of the joint Shannon entropy of  $X$  and  $Y$. This ratio quantifies the proportion of shared information relative to the total information. It exhibits non-negativity and symmetry, and satisfies  $\bar{I}(X;Y)=0$  if and only if  $X$  and  $Y$  are statistically independent.

The method of identifying chaos using mutual information involves tracking the trajectories of two particles released from proximate initial positions ($\Delta r=10^{-8}$).
The Shannon entropy of each particle and their joint Shannon entropy are computed separately, after which their mutual information is calculated using Eq. (43) and normalized.
This methodology is termed MIPP \cite{Cao:2024rvo}.
In chaotic systems, due to exponential sensitivity to initial conditions, the trajectories of the two particles rapidly diverge over time.
Consequently, their statistical correlation diminishes in the long term, causing the mutual information $I(X;Y)$ and the MIPP value converges to 0.
In regular  systems, trajectories maintain strong correlation, mutual information  $I(X;Y)$  remains significant, and the MIPP value approaches 1.
This method exhibits high sensitivity to initial value perturbations and is particularly suited for capturing transitions between chaotic and regular regimes.

\subsection{Parameter space scanning and chaos detection results}
Using the three aforementioned chaos indicators, we analyze the influence of particle energy  $E$, angular momentum  $L$, parameter $e^{-\nu}$ and magnetic charge  $Q_{m}$  on orbital chaos characteristics, and investigate the modulation effects of these parameters.

\begin{figure}[H]
\centering
\includegraphics[width=0.4\textwidth]{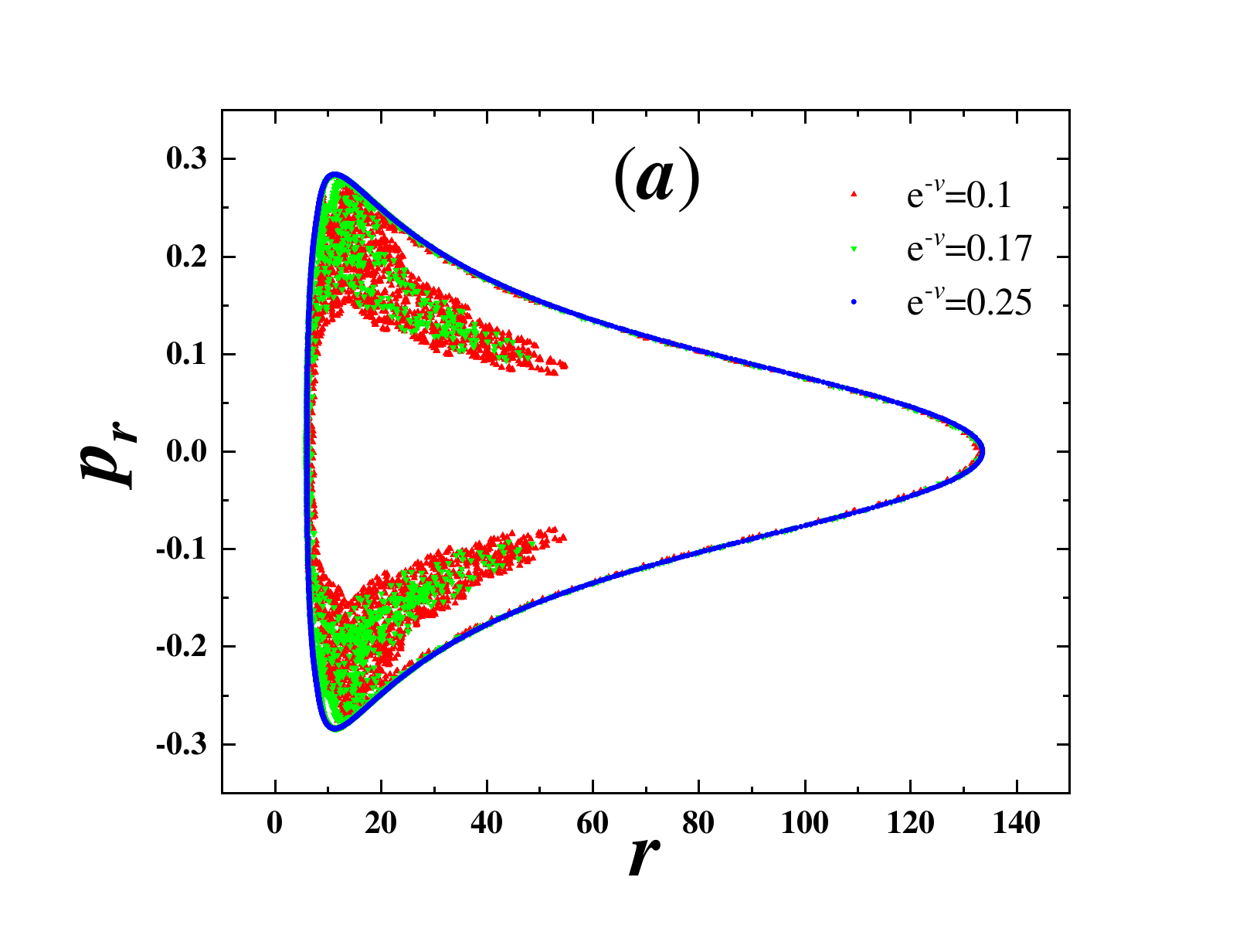}
\includegraphics[width=0.4\textwidth]{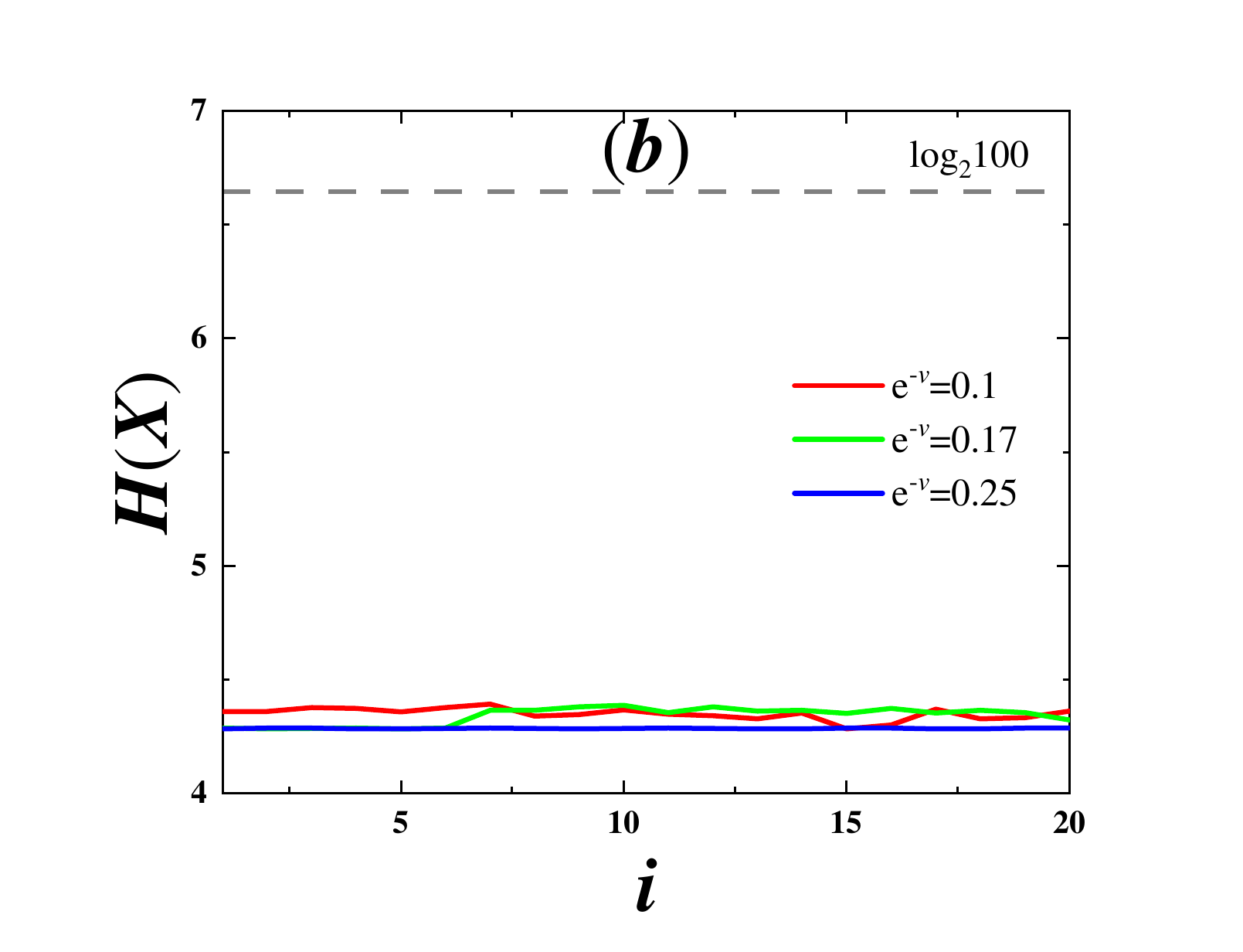}
\includegraphics[width=0.4\textwidth]{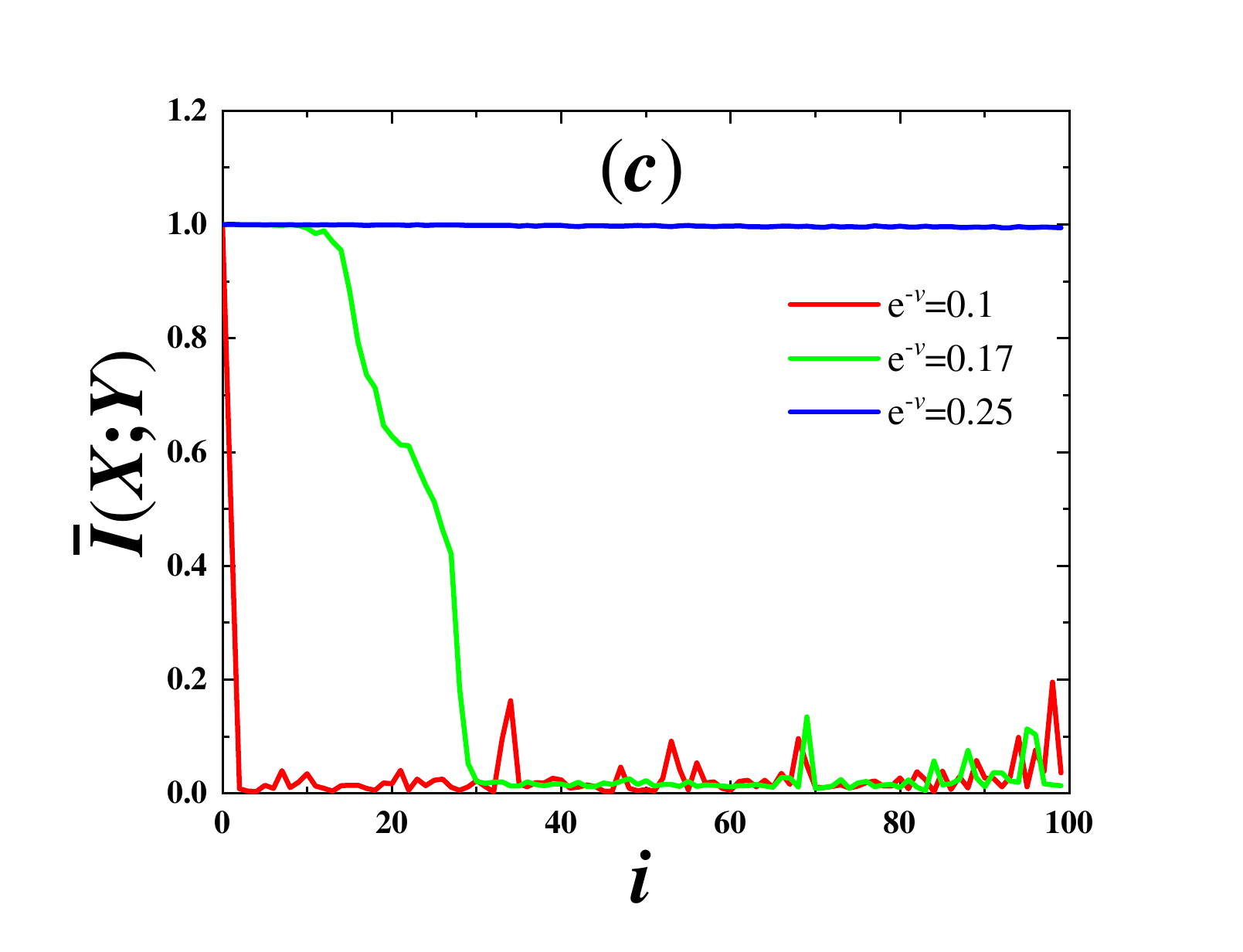}
\includegraphics[width=0.4\textwidth]{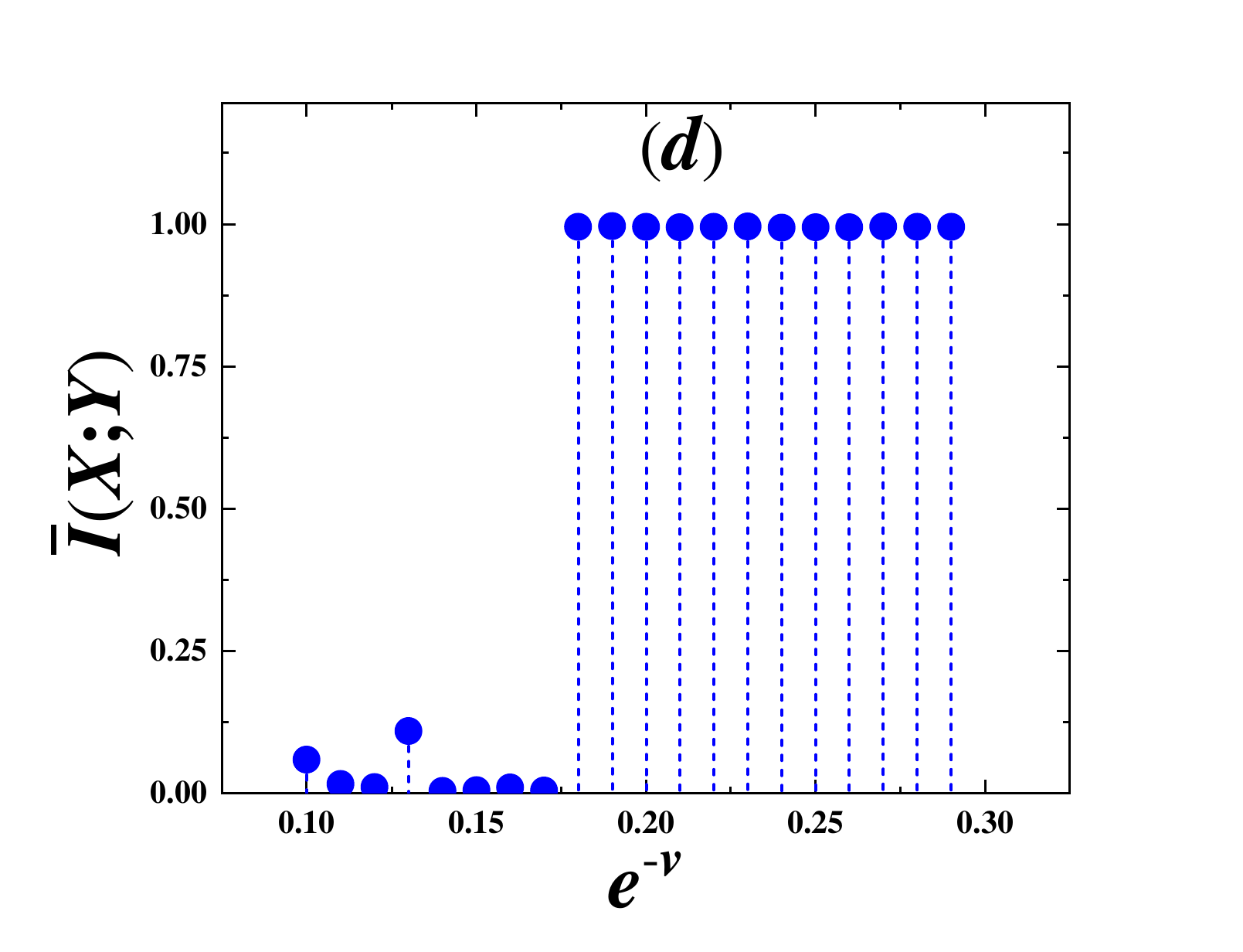}
\caption{Chaos indicators for different orbits with $E=0.996$, $L=4.1$, $Q_{m}=1$, $\beta=-0.0008$, $r=6$, $\theta=\frac{\pi}{2}$, $p_{r}=0$, and integration step $h=1$.
(a), (b), and (c) display the Poincar$\acute{e}$ section, Shannon entropy, and MIPP evolution for $e^{-\nu}=0.1$, $0.17$, and $0.25$, respectively.
(d) shows the final MIPP values at the 100th interval for a sequence of $e^{-\nu}$ values, where each blue point represents the MIPP result for a specific $e^{-\nu}$ sampled in the parameter space.}\label{fig3}
\end{figure}

In Fig. 3(a), when  $e^{-\nu}$  is small (e.g.,  $e^{-\nu}=0.1$), the intersection points on the Poincar$\acute{e}$ section are randomly distributed, indicating chaotic motion.
When  $e^{-\nu}$  is large (e.g.,  $e^{-\nu}=0.25$), the intersection points form closed curves, indicating regular motion.
The state of these three orbits can be identified via Shannon entropy.
In Fig. 3(b), for  $e^{-\nu}=0.1$, the orbital Shannon entropy increases and exhibits significant fluctuations, whereas for  $e^{-\nu}=0.25$, the orbital Shannon entropy decreases with its fluctuation amplitude diminishing toward 0, appearing nearly linear.
This is consistent with the Poincar$\acute{e}$ section results. MIPP demonstrates high sensitivity for chaos detection.
In Fig. 3(c), for  $e^{-\nu}=0.1$, the MIPP values across 100 intervals rapidly decrease to 0 and fluctuate nearby, while for  $e^{-\nu}=0.25$, the MIPP values stabilize and fluctuate near 1.
The MIPP at the 100th interval represents the system's final state.
Fig. 3(d) reveals the impact of varying  $e^{-\nu}$  on orbital dynamics: $e^{-\nu}=0.17$   serves as the critical threshold for the chaos to regular motion.
Specifically, the system exhibits chaos when  $e^{-\nu}\leq0.17$, and regular motion when $e^{-\nu}>0.17$.

\begin{figure}[H]
\centering
\includegraphics[width=0.4\textwidth]{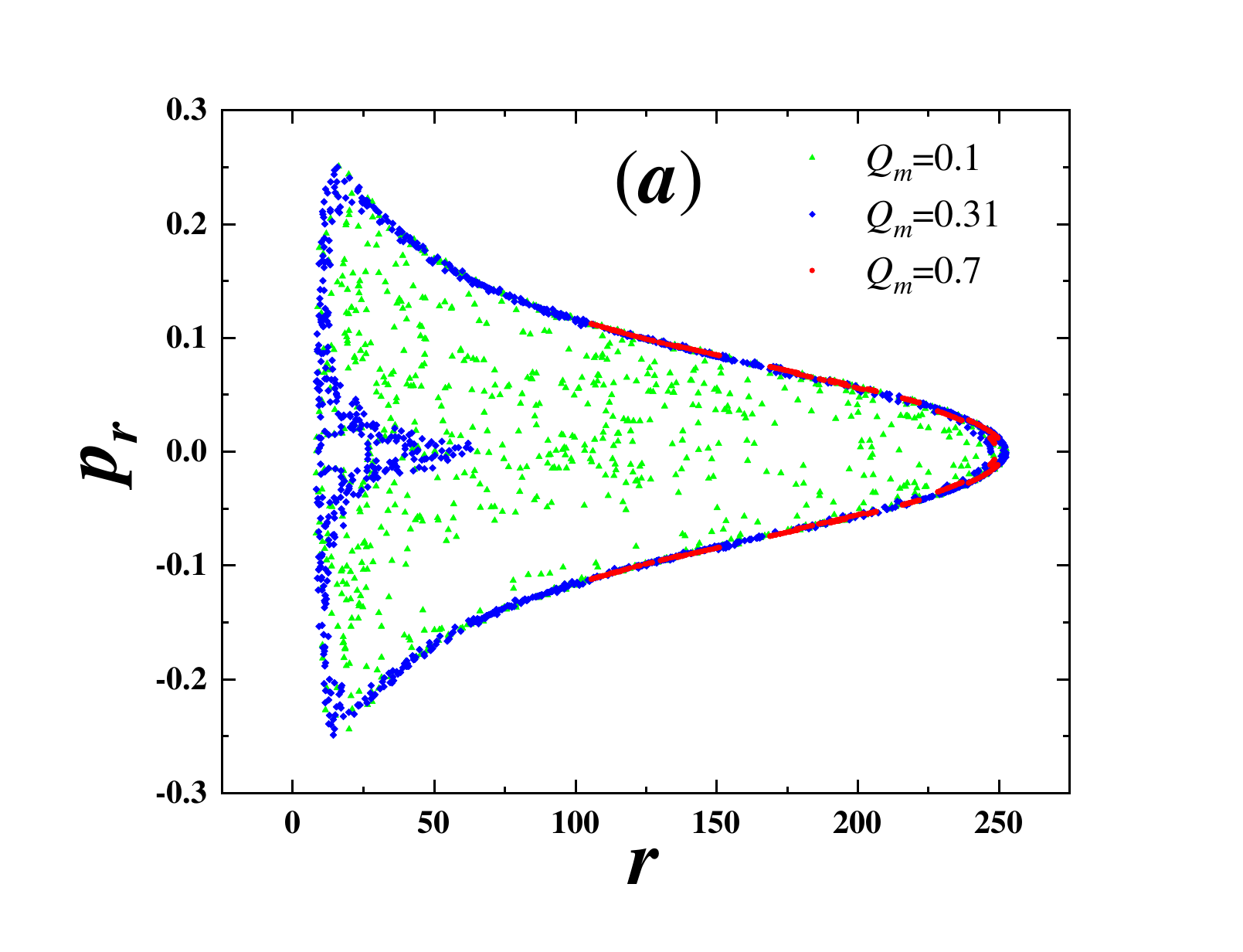}
\includegraphics[width=0.4\textwidth]{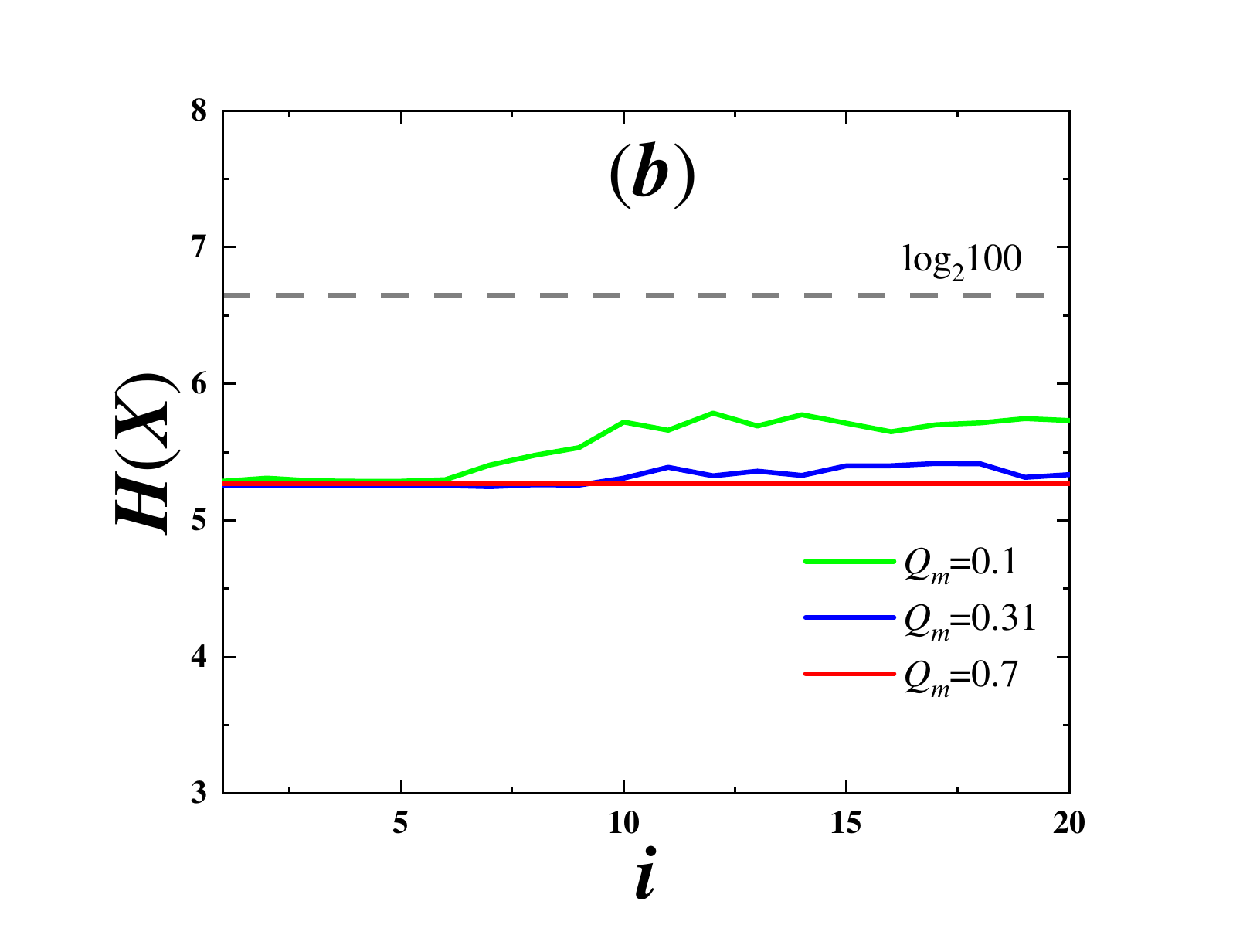}
\includegraphics[width=0.4\textwidth]{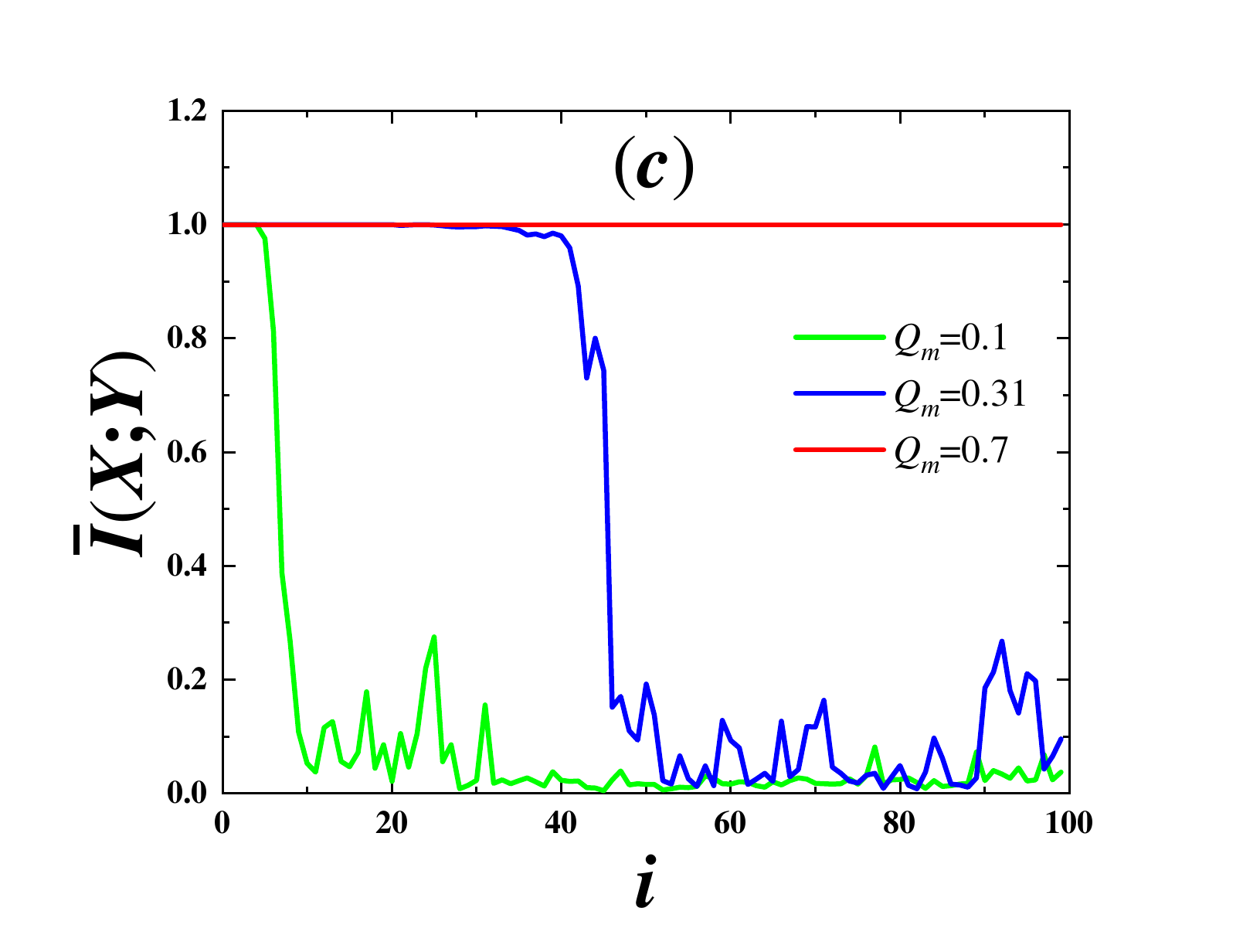}
\includegraphics[width=0.4\textwidth]{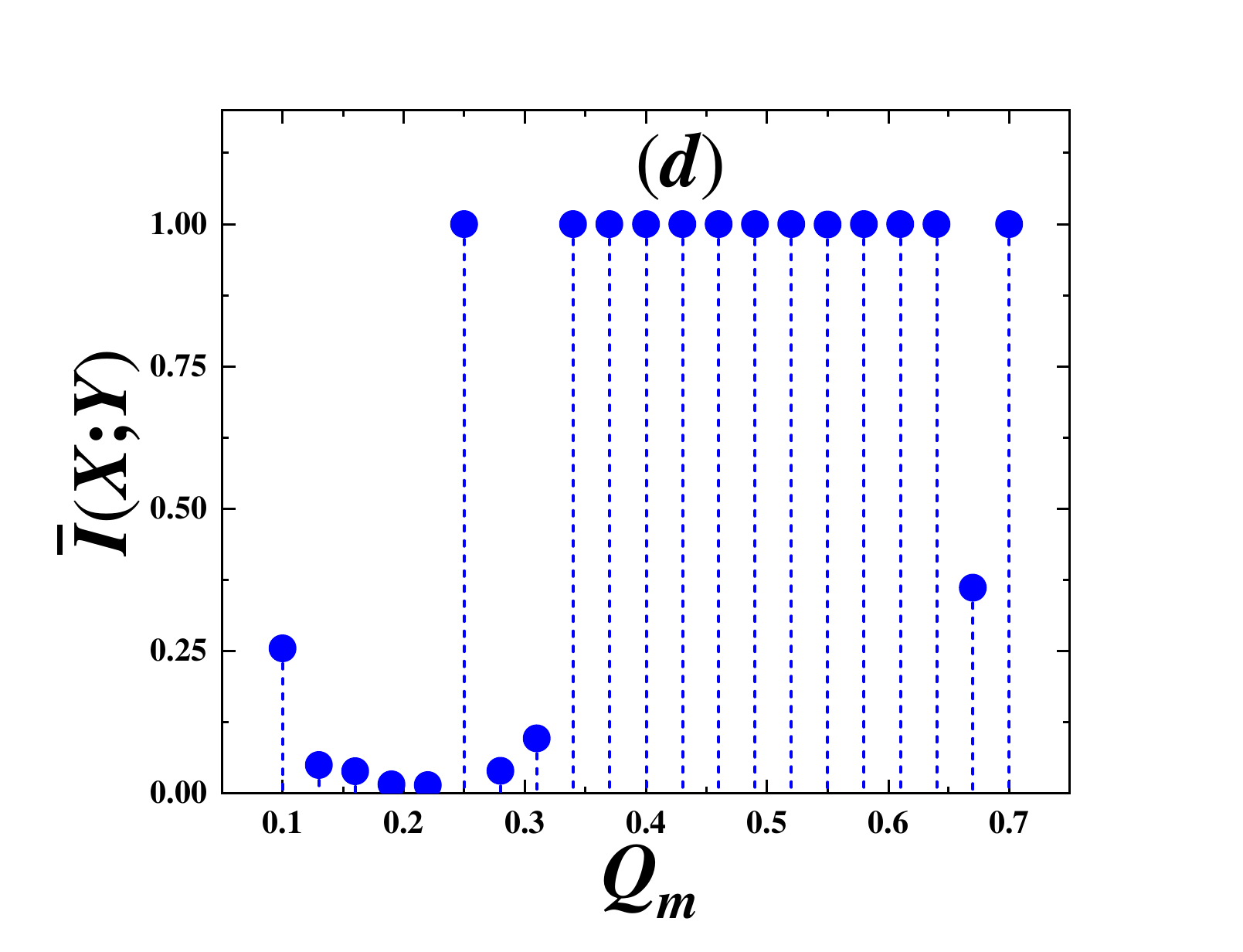}
\caption{Chaos indicators for different orbits with $E=0.997$, $L=4.6$, $e^{-\nu}=1$, $\beta=0.00049$, $r=30$, $\theta=\frac{\pi}{2}$, $p_{r}=0$, and integration step $h=1$.
(a), (b), and (c) display the Poincar$\acute{e}$ section, Shannon entropy, and MIPP evolution for $Q_{m}=0.1$, $0.31$, and $0.7$, respectively.
(d) shows the final MIPP values at the 100th interval for a sequence of $Q_{m}$ values, where each blue point represents the MIPP result for a specific $Q_{m}$ sampled in the parameter space.
}\label{fig4}
\end{figure}

The conclusions from Fig. 4(a)-(c) are consistent with those of Fig. 3.
However, in Fig. 4(d), the influence of varying  $Q_{m}$  on orbital dynamics exhibits greater complexity. When $Q_{m}\leq0.31$, the orbits exhibit widespread chaotic states.
Notably, at $Q_{m}=0.25$ the orbit is regular, whereas at $Q_{m}=0.67$  it is chaotic.

\begin{figure}[H]
\centering
\includegraphics[width=0.32\textwidth]{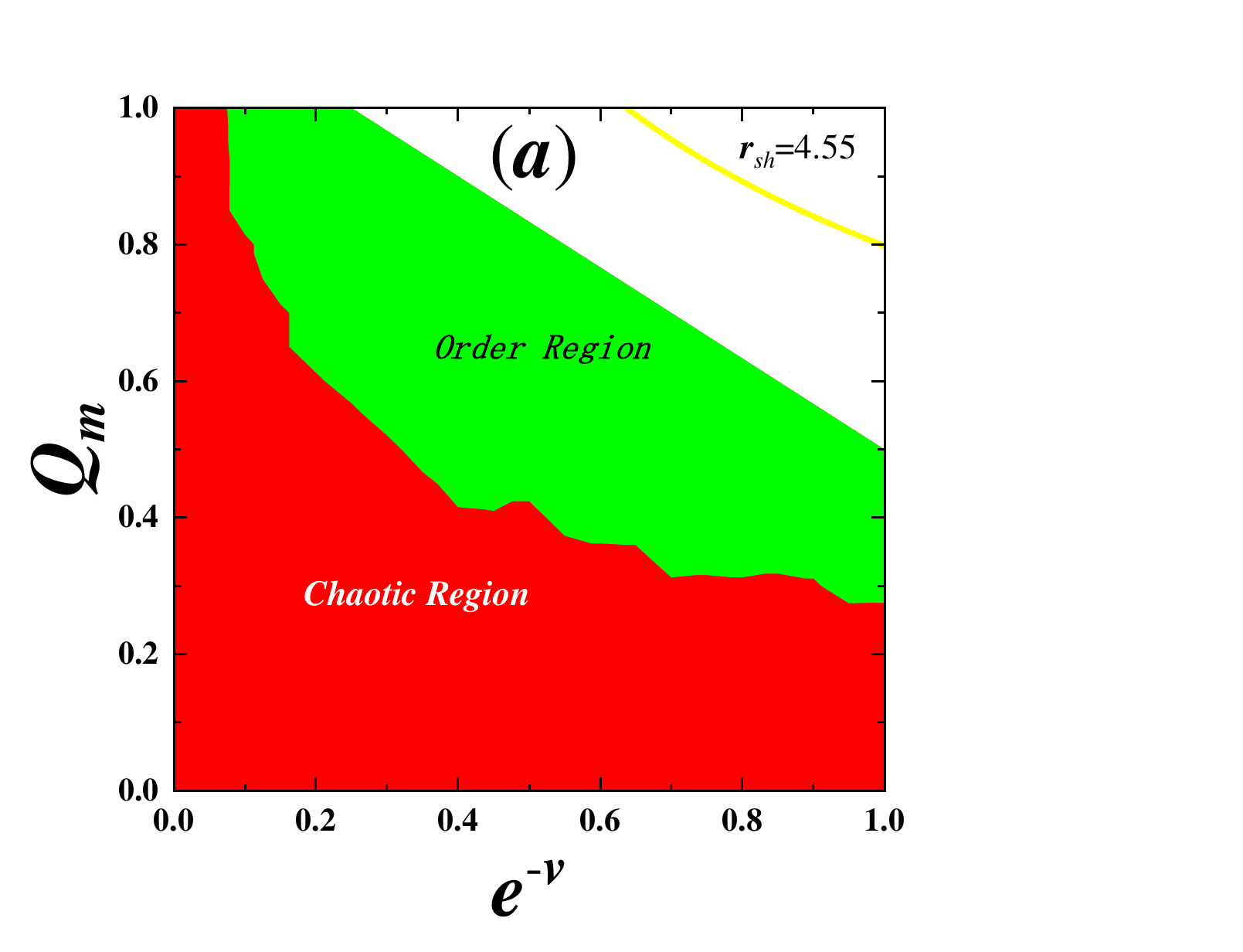}
\includegraphics[width=0.32\textwidth]{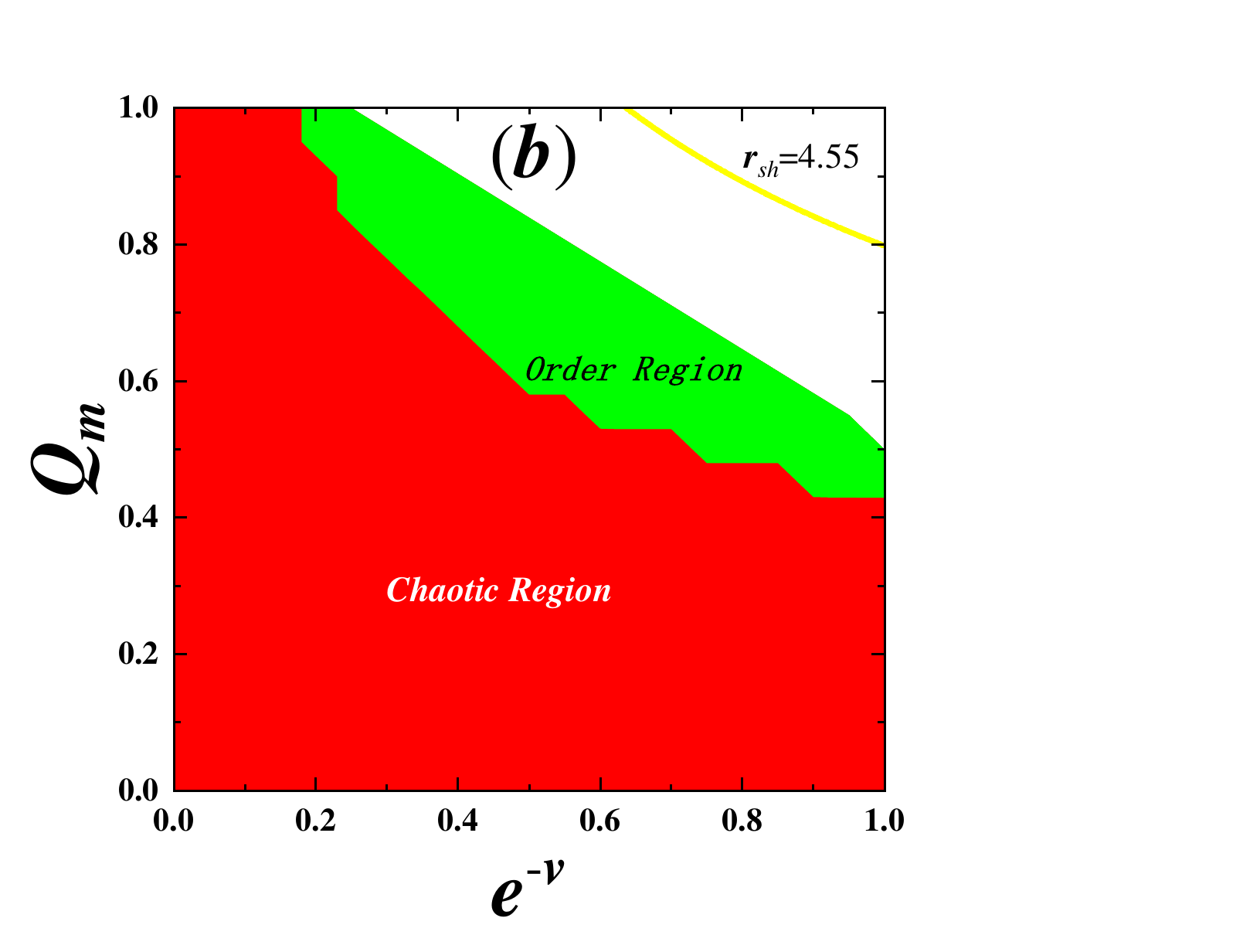}
\includegraphics[width=0.32\textwidth]{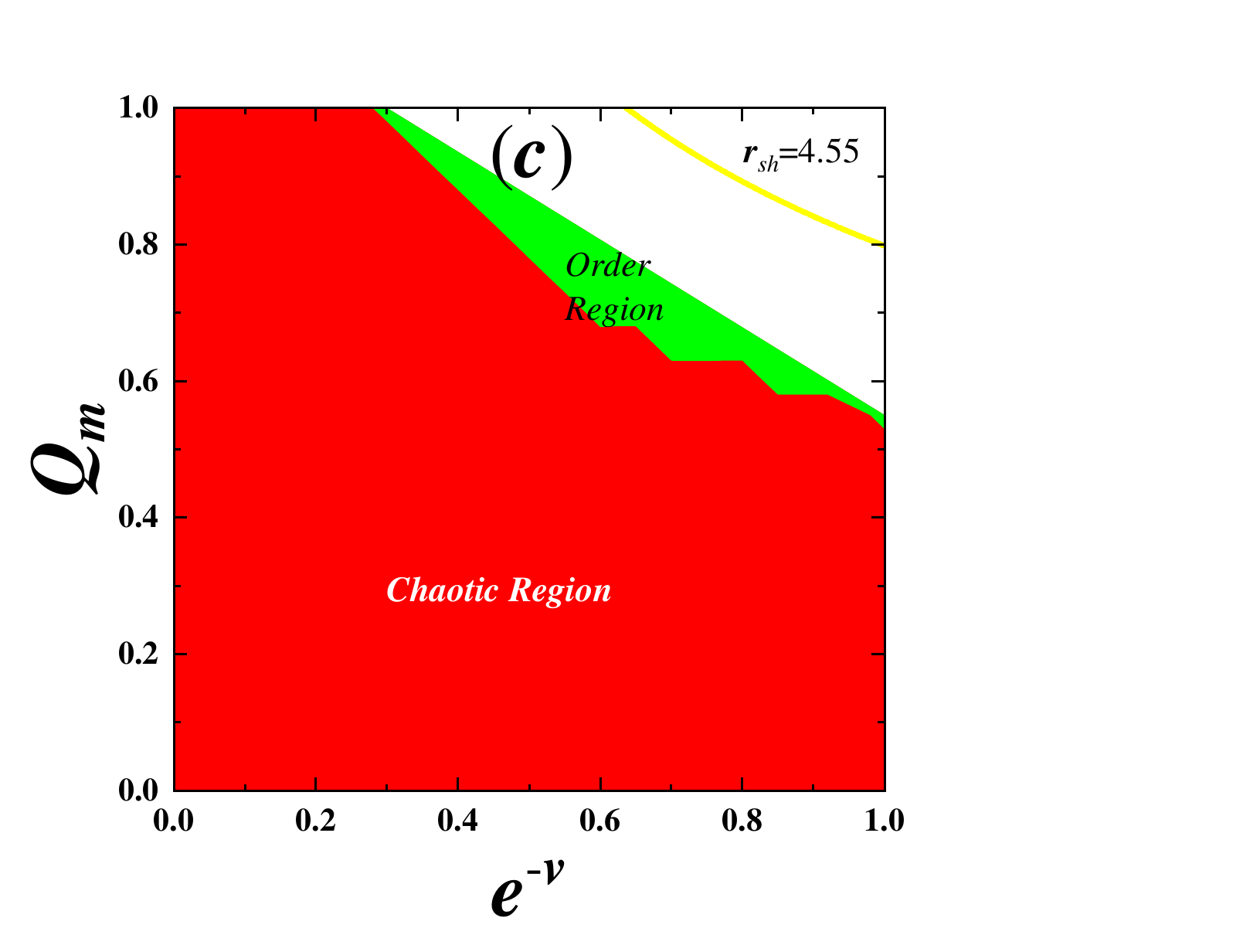}
\caption{Two-dimensional MIPP parameter space scans. The orbital parameters are $L=4.1$,  $\beta=-0.0008$, $r=6$, $\theta=\frac{\pi}{2}$, $p_{r}=0$, and integration step $h=1$.
The red and green regions represent chaotic and regular motion, respectively, while the yellow curve indicates the critical shadow radius constraint.
(a)-(c): Scan plots for increasing energy $E=0.9955$, $0.996$, and $0.997$, respectively.}\label{fig5}
\end{figure}

In Fig. 5, chaotic regions are concentrated in the lower-left area, while regular regions dominate the upper-right section.
Energy variation significantly impacts orbital states.
At  $E=0.9955$, order regions has the maximum proportion.
When  $E=0.996$, regular regions diminish while chaotic regions expand.
At  $E=0.997$, regular regions almost vanish, yielding maximum chaotic coverage.
The boundary between chaotic and regular regions progressively contracts with increasing energy, aligning with the enhanced chaotic nature of orbits at higher energies \cite{Kopacek:2010yr}.

\begin{figure}[H]
\centering
\includegraphics[width=0.32\textwidth]{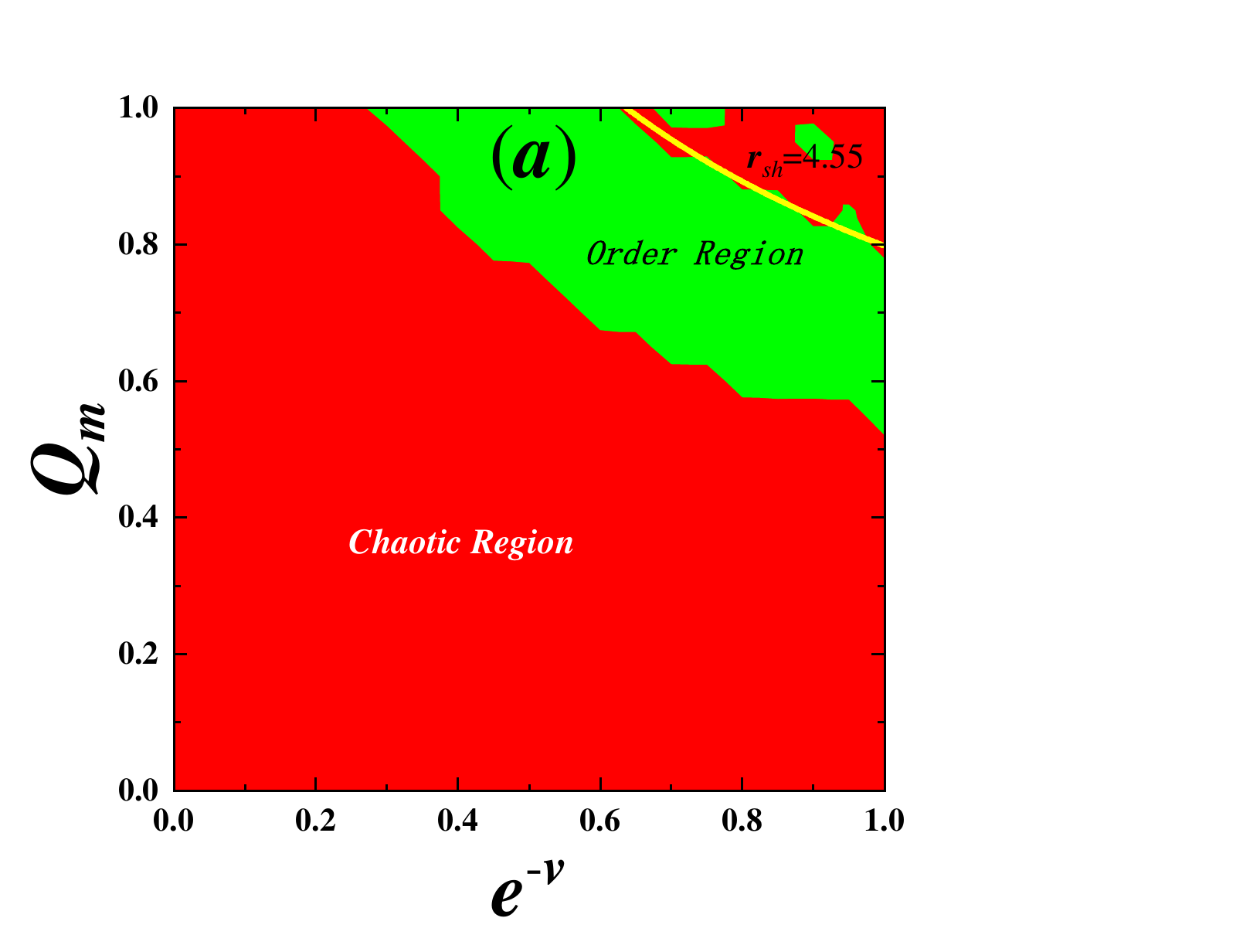}
\includegraphics[width=0.32\textwidth]{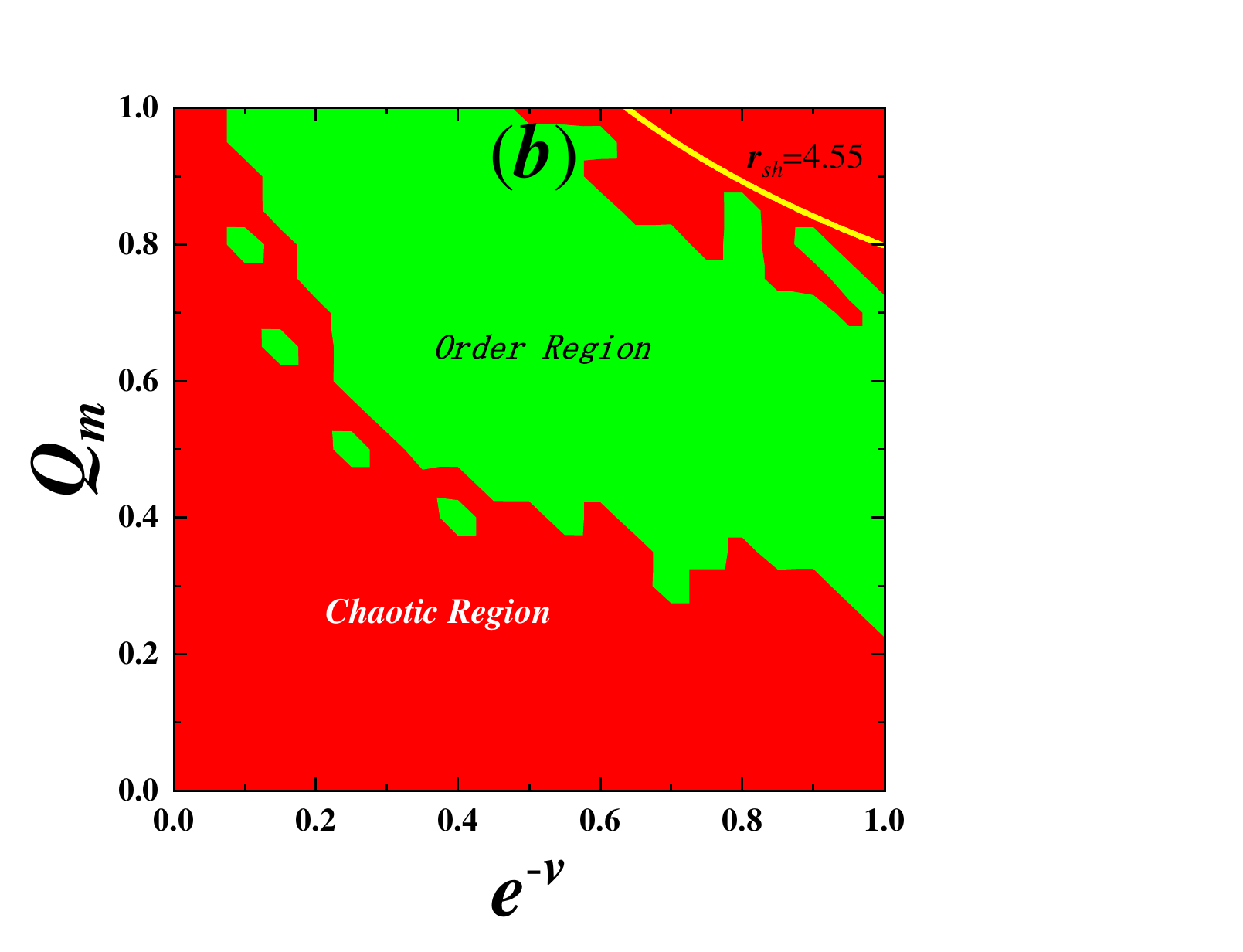}
\includegraphics[width=0.32\textwidth]{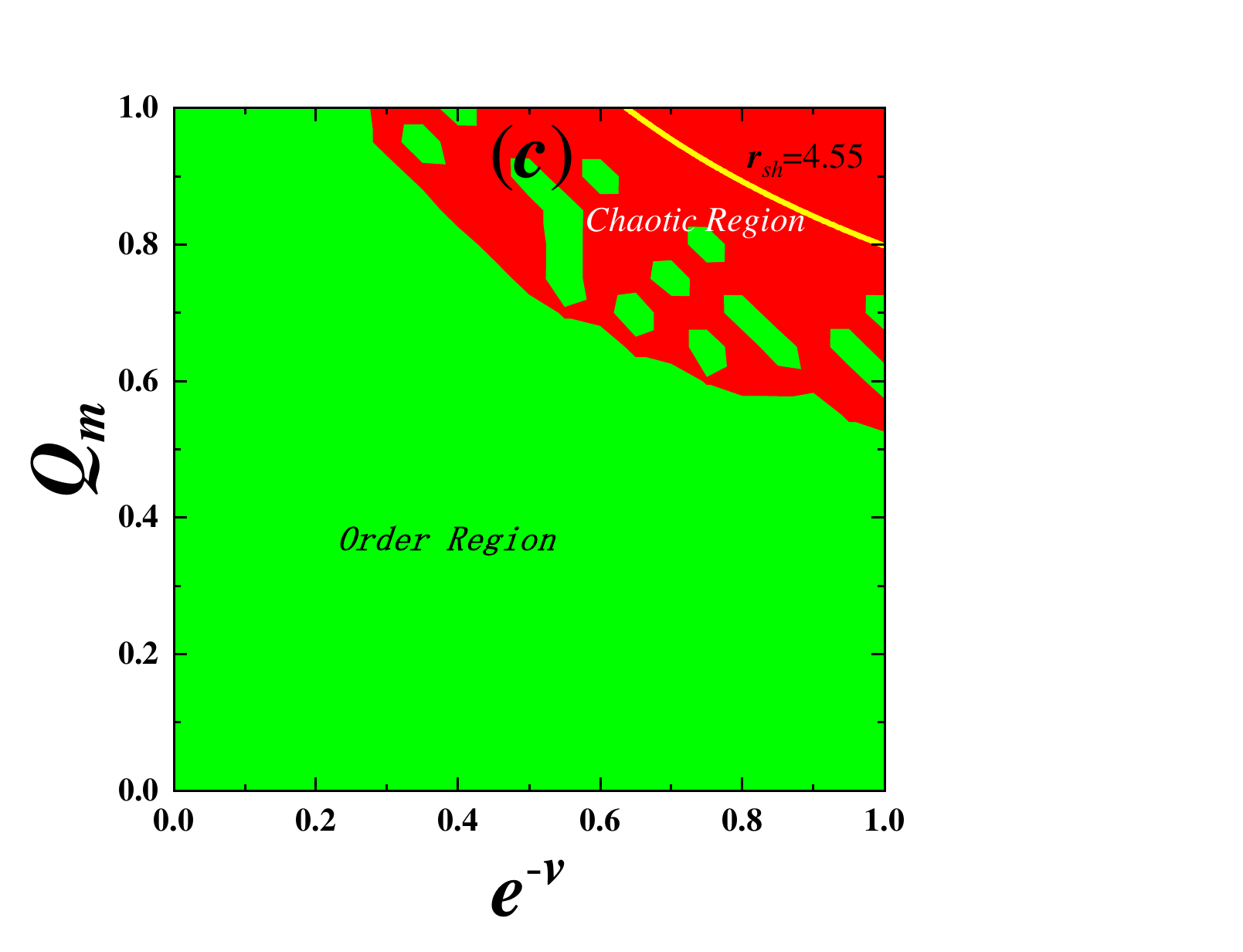}
\caption{Two-dimensional MIPP parameter space scans. The orbital parameters are $E=0.997$,  $\beta=0.00049$, $r=30$, $\theta=\frac{\pi}{2}$, $p_{r}=0$, and integration step $h=1$.
The red and green regions represent chaotic and regular motion, respectively, while the yellow curve indicates the critical shadow radius constraint.
(a)-(c): Scan plots for increasing angular momentum $L=4.5$, $4.6$, and $4.7$, respectively.}\label{fig6}
\end{figure}

The effect of angular momentum on the orbital state is relatively complex. In Fig. 6, as the angular momentum gradually increases, the chaotic region in the lower-left decreases, while the chaotic region in the upper-right increases. Meanwhile, the regular region shifts towards the lower-left, and multiple boundary lines between the chaotic and regular regions appear.

\begin{figure}[H]
\centering
\includegraphics[width=0.32\textwidth]{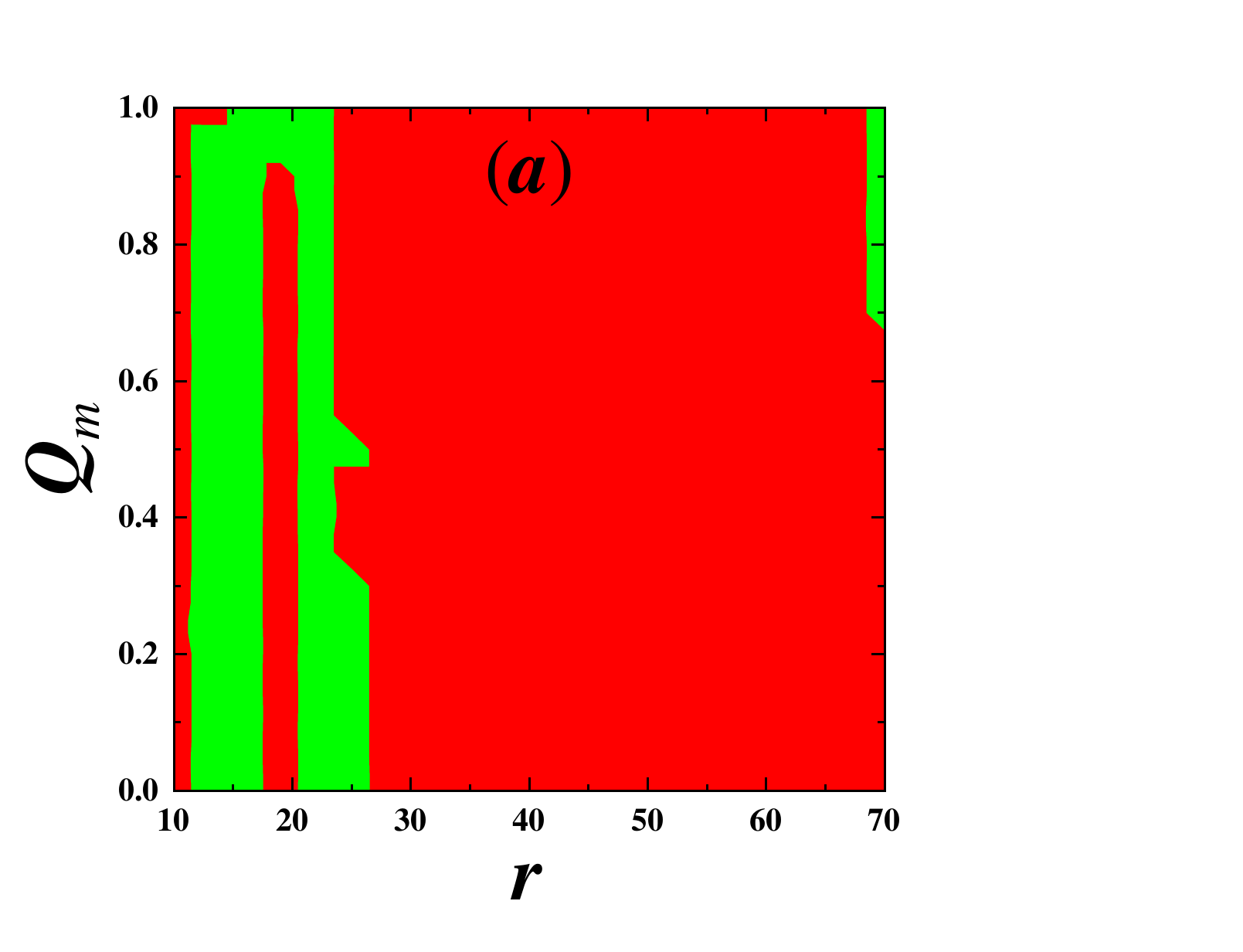}
\includegraphics[width=0.32\textwidth]{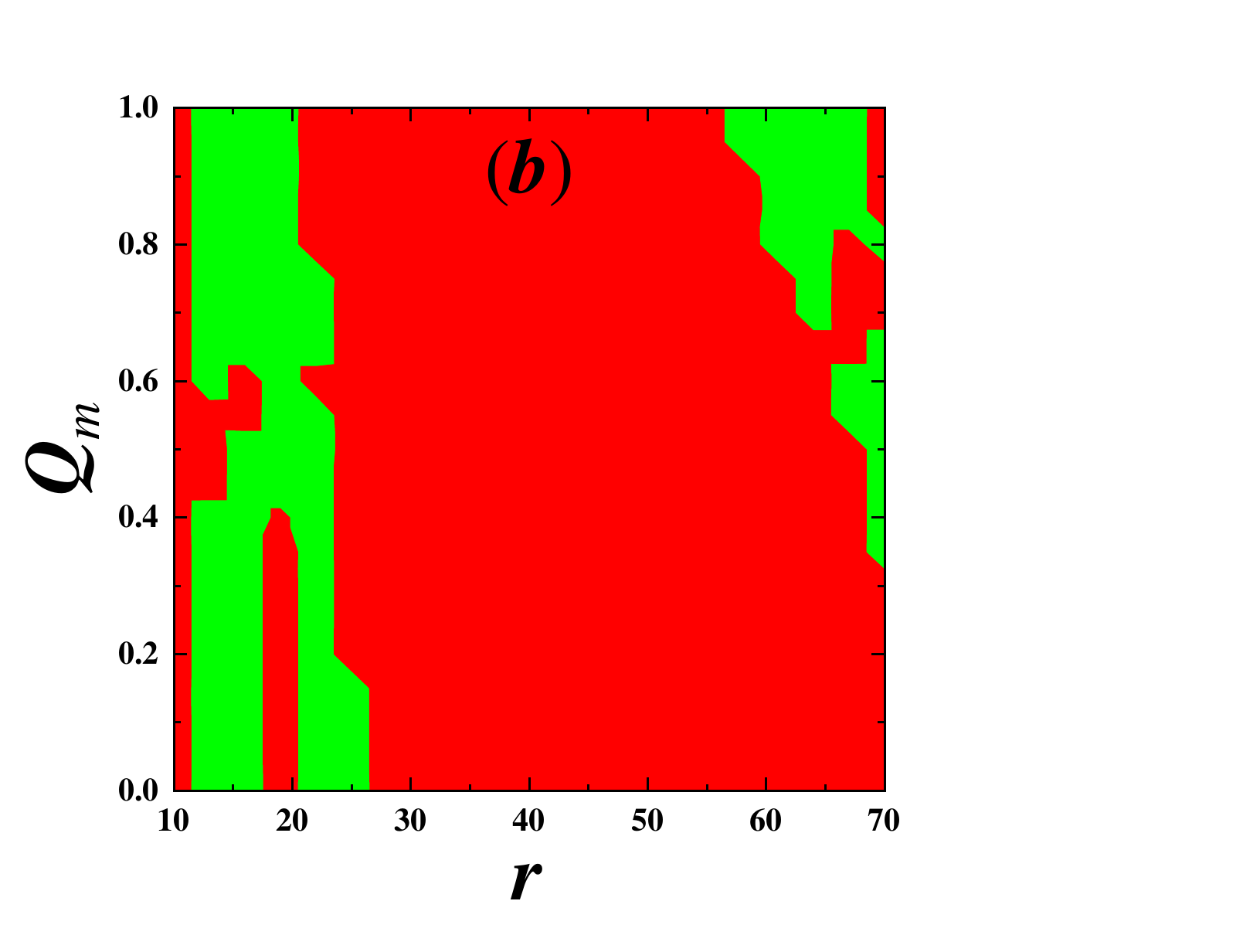}
\includegraphics[width=0.32\textwidth]{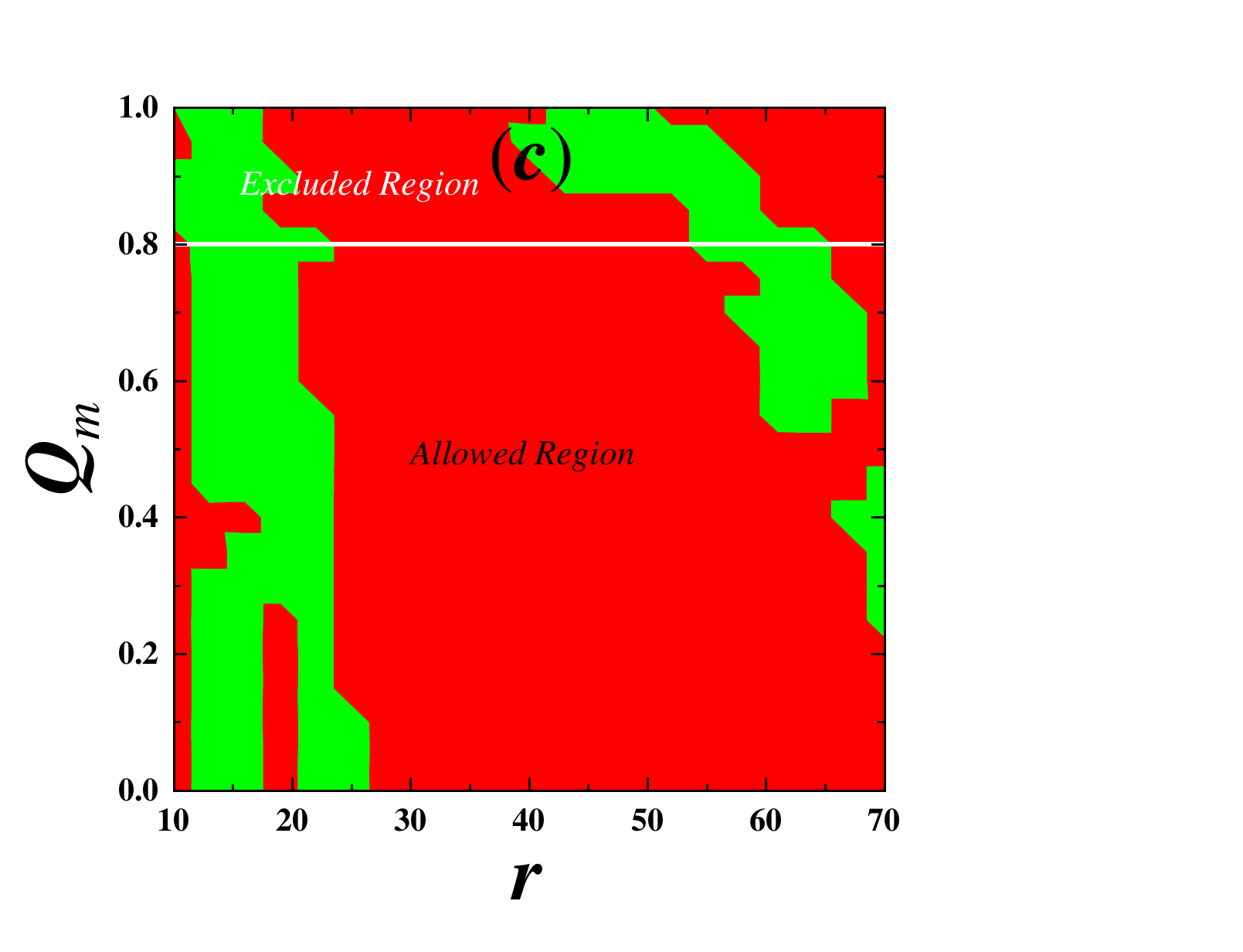}
\caption{Two-dimensional MIPP parameter space scans. The orbital parameters are $E=0.995$, $L=4.6$, $\beta=0.001$, $\theta=\frac{\pi}{2}$, $p_{r}=0$, and integration step $h=1$.
The red and green regions represent chaotic and regular motion, respectively.
(a)-(c): Scan plots for increasing $e^{-\nu}=0.1$, $0.5$, and $1$, respectively.}\label{fig7}
\end{figure}

In Figs. 7 and 8, the influence of varying  $e^{-\nu}$  and  $Q_{m}$  on orbital dynamics exhibits similarities.
Regular regions persistently exist near the black hole event horizon.
As  $e^{-\nu}$  or  $Q_{m}$  increases, the regular regions expand, and the right side regular zones  converge toward the event horizon.
When combined with Figs. 5 and 6, it is evident that the impact of parameters  $e^{-\nu}$  or  $Q_{m}$  on orbital states is less pronounced than that of  $E$   and  $L$.

\begin{figure}[H]
\centering
\includegraphics[width=0.32\textwidth]{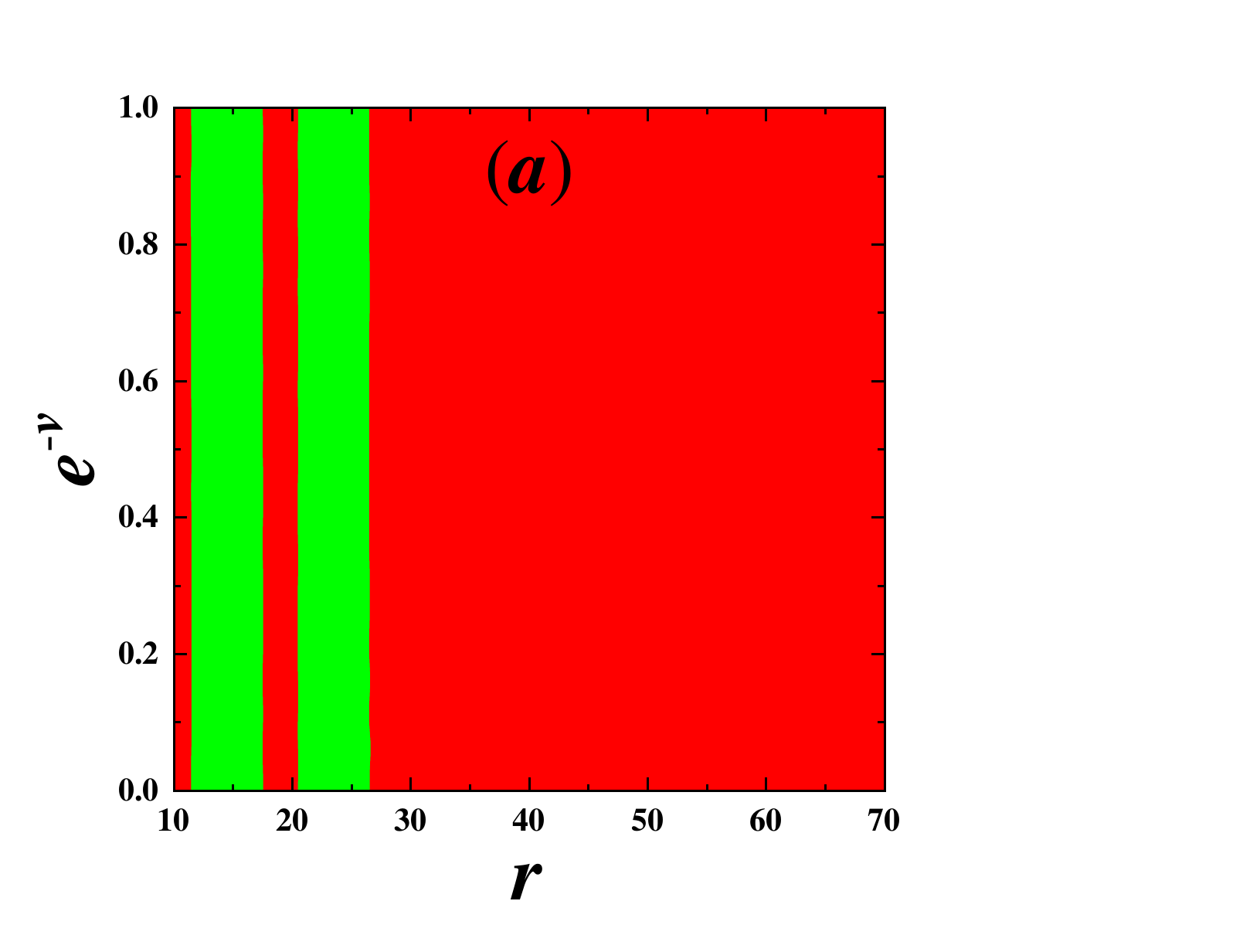}
\includegraphics[width=0.32\textwidth]{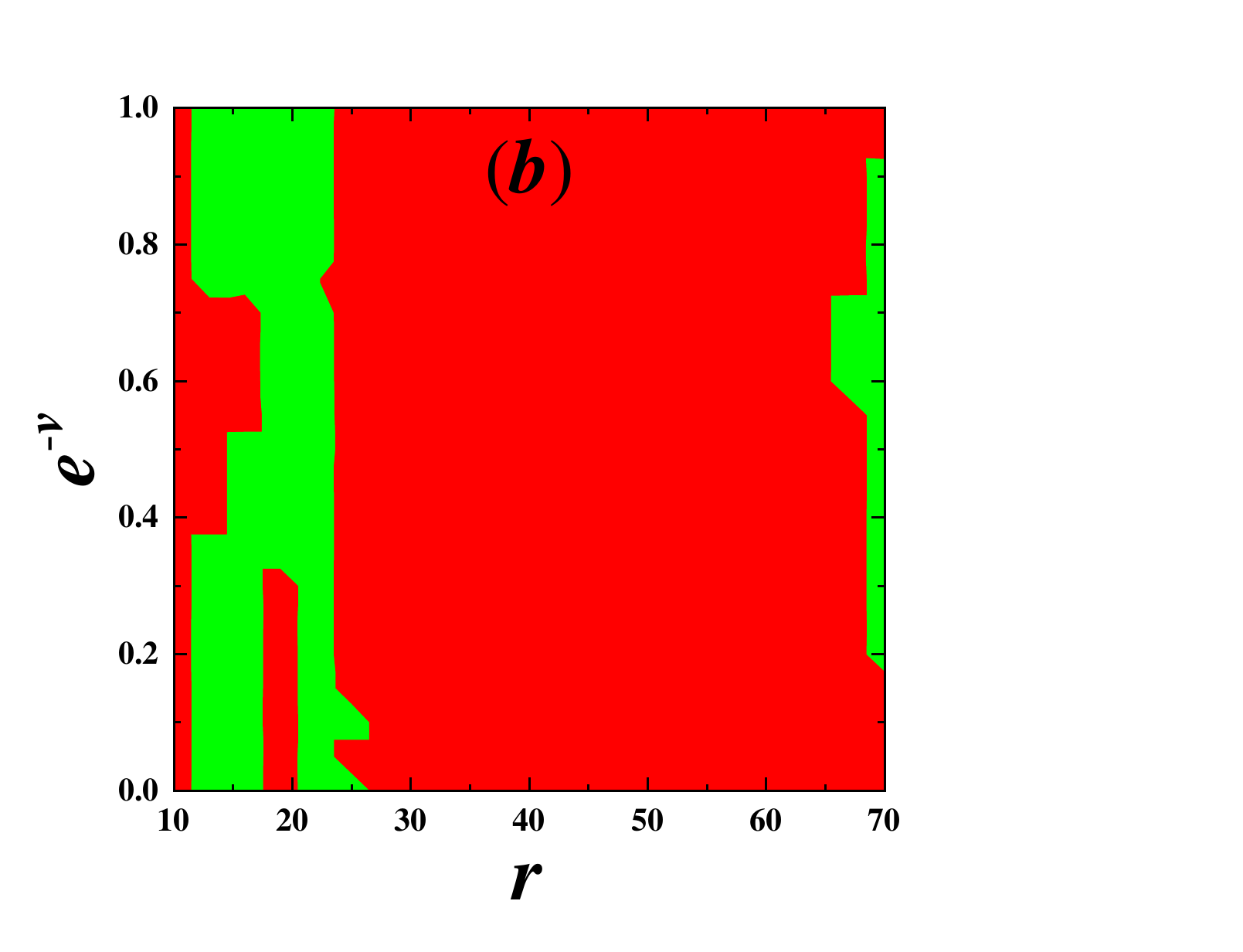}
\includegraphics[width=0.32\textwidth]{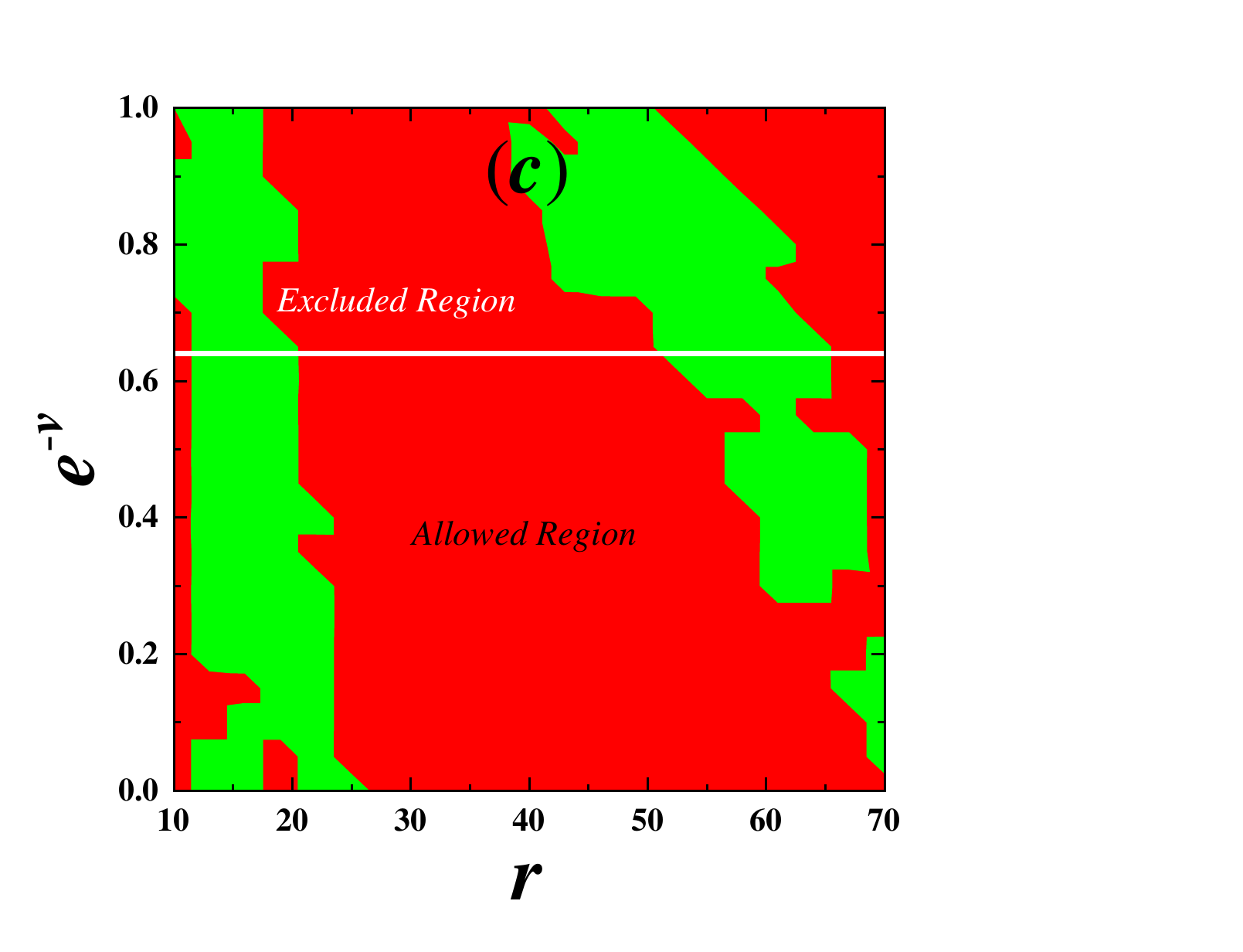}
\caption{Two-dimensional MIPP parameter space scans. The orbital parameters are $E=0.995$, $L=4.6$, $\beta=0.001$, $\theta=\frac{\pi}{2}$, $p_{r}=0$, and integration step $h=1$.
The red and green regions represent chaotic and regular motion, respectively.
(a)-(c): Scan plots for increasing $Q_{m}=0.1$, $0.5$, and $1$, respectively.}\label{fig8}
\end{figure}

In summary, the particle energy $E$ plays a dominant role in enhancing chaotic dynamics by enlarging the accessible region of the global phase space.
By contrast, an increase in the particle angular momentum $L$ tends to suppress chaotic motion.
The gravitational parameters $e^{-\nu}$ and $Q_{m}$ introduce comparatively weaker modifications to the global phase-space structure than the dynamical parameters $E$ and $L$. 
To provide a physical interpretation for these results, we analyze the dominant terms of the Hamiltonian extracted from Eq. (16), which include the following principal contributions:

\begin{eqnarray}
K_{1}=-\frac{E^{2}+L\beta}{2}-\frac{E^{2}}{r}+\frac{E^{2}}{2r^{2}}e^{-\nu}Q_{m}^{2}+\frac{\beta^{2}}{8}r^{2}\sin^{2}\theta+\frac{L^{2}}{2r^{2}\sin^{2}\theta}+...,
\end{eqnarray}
the expression is considered when $\frac{2}{r}\ll 1$ and $\frac{e^{-\nu}Q_{m}^{2}}{r^{2}}\ll 1$.

The second term in Eq. (45) corresponding to the second term of Eq. (2) acts as the black hole gravity to the particle.
However, $e^{-\nu}$ and $Q_{m}$ in the third term of Eq. (45) corresponding to the third
term of Eq. (2) gives a gravitational repulsive force contribution to the particle.
The gravitational force from $E$ in the second term of Eq. (45) is more important than the gravitational repulsive force from $e^{-\nu}$ and $Q_{m}$ in the third term of Eq. (45).
The fourth term in Eq. (45) corresponds to a magnetic field force as a gravitational effect.
The fifth term corresponds to an inertial centrifugal force from the particle angular momentum.
When $E$ increases, the effect of gravity becomes stronger, and the motion of the particle changes accordingly.
As a result, chaos would become stronger from the global phase space structures when chaos can occur.
On the contrary, an increase of the angular momentum leads to enlarging the repulsive force effects; equivalently, it weakens the gravitational effects and decreases the strength of chaos.
In contrast, the parameters $e^{-\nu}$ and $Q_{m}$ mainly modify the background spacetime geometry through the metric function of the Einstein-ModMax black hole.
Within the parameter range considered in this work, these parameters only introduce moderate corrections to the spacetime structure.
Consequently, their influence on the global phase-space structure of the particle motion is weaker than that of the conserved quantities $E$ and $L$.

\section{Conclusion}\label{sec:four}

In this paper, we analyze the chaotic dynamics of charged particles in the Einstein-ModMax weakly magnetized black hole spacetime using Shannon entropy and MIPP to investigate the influence of different parameters on the chaotic and regular regions of orbital motion. Our results show that the particle energy $E$ plays a dominant role in determining the dynamical state of the orbit: as $E$ increases, the chaotic region gradually enlarges while the regular region shrinks. The angular momentum $L$ has a more complex influence; with increasing $L$, the chaotic region tends to decrease and the regular region expands, while multiple transition boundaries may appear. In contrast, the parameters $e^{-\nu}$ and $Q_{m}$ have comparatively weaker impacts on the orbital state.

Shannon entropy and MIPP provide  effective quantitative measures for characterizing the complexity of orbital dynamics, which we refer to as a descriptive measure of orbital complexity. Shannon entropy can distinguish different dynamical regimes: as the orbit evolves from periodic to
quasi-periodic and eventually to chaotic motion, the entropy generally increases and exhibits stronger fluctuations for chaotic trajectories. MIPP characterizes the statistical correlation between nearby particle trajectories. In chaotic regimes, the correlation rapidly decreases, causing MIPP to approach zero, whereas regular motion maintains stronger correlations and leads to larger MIPP values.
Consequently, these tools are particularly well-suited for identifying global phase-space structures in complex spacetimes, providing a promising methodology for further exploration of orbital dynamics in alternative gravity theories.

\noindent \textbf{Acknowledgements}

\noindent \textbf{Data availability statement} Data will be made available on reasonable request.

\noindent \textbf{Declarations}

\noindent \textbf{Conflict of interest}
The authors declared that there is no conflict of interest in this manuscript.

\end{document}